\documentclass{jfm}

\usepackage{graphicx}
\usepackage[percent]{overpic}
\usepackage{epstopdf, epsfig}
\usepackage{graphicx}
\usepackage{dcolumn}
\usepackage{bm}
\usepackage{color}
\usepackage{array}
\usepackage{amsmath,amssymb,stmaryrd} 
\usepackage{hyperref}
\usepackage{soul} 
\definecolor{blue3}{rgb}{0, 0.1770, 0.3410}

\hypersetup{
	colorlinks=true,
	linktoc=all,
	linkcolor=blue,
	citecolor=blue,
}

\definecolor{blue2}{rgb}{0, 0.4470, 0.7410}
\definecolor{red2}{rgb}{0.8500, 0.1250, 0.0480} 
\definecolor{orange2}{rgb}{0.8500, 0.3250, 0.0980} 
\definecolor{yellow2}{rgb}{0.9290, 0.6940, 0.1250}
\definecolor{purple2}{rgb}{0.4940, 0.1840, 0.5560}
\definecolor{green2}{rgb}{0.4660, 0.6740, 0.1880}
\definecolor{ltblue2}{rgb}{0.3010, 0.7450, 0.9330}
\definecolor{dkred2}{rgb}{0.6350, 0.0780, 0.1840}
\definecolor{gray2}{rgb}{0.22, 0.22, 0.3}
\definecolor{gray3}{rgb}{0.7, 0.7, 0.7}

\shorttitle{Thermal forcing on shear layers}
\shortauthor{C.-A. Yeh, P. M. Munday and K. Taira}

\title{Laminar free shear layer modification using localized periodic heating}

\author{Chi-An Yeh\aff{1},
		Phillip M. Munday\aff{1} \and
		Kunihiko Taira\aff{1}
		\corresp{\email{cy13d@my.fsu.edu}}}

\affiliation{\aff{1}Department of Mechanical Engineering, Florida State University,
Tallahassee, FL 32310, USA}

\begin{document}

\maketitle

\begin{abstract}
The application of local periodic heating for controlling a spatially developing shear layer downstream of a finite-thickness splitter plate is examined by numerically solving the two-dimensional Navier-Stokes equations. At the trailing edge of the plate, oscillatory heat flux boundary condition is prescribed as the thermal forcing input to the shear layer.  The thermal forcing introduces low level of oscillatory surface vorticity flux and baroclinic vorticity at the actuation frequency in the vicinity of the trailing edge.  The produced vortical perturbations can independently excite the fundamental instability that accounts for shear layer roll-up as well as the subharmonic instability that encourages the vortex pairing process farther downstream.  We demonstrate that the nonlinear dynamics of a spatially developing shear layer can be modified by local oscillatory heat flux as a control input.  We believe that this study provides a basic foundation for flow control using thermal-energy-deposition-based actuators such as thermophones and plasma actuators.
\end{abstract}

\begin{keywords}
shear layer, flow control, instability.
\end{keywords}

\section{Introduction}
Shear layer may be the most common flow that arises in virtually every applications, including jets, flow over a cavity, separated boundary layer over an aerodynamic body, and merging of two streams behind the trailing edge of an airfoil.  Accordingly, active control of shear layer flows has been a major area of interest for the community.  Early efforts on control of shear layers arising from separated flows have been summarized by \citet{Gad-el-Hak:JFE1991}.  For jet flows, \citet{Wiltse:PF98} examined the use of dissipative-range forcing with piezoelectric actuators and found that the high-frequency actuation is able to enhance the turbulent energy cascade and increase the dissipation of turbulent kinetic energy.  Later, \citet{SeifertPack:AIAAJ1999} examined the effect of synthetic jet actuation for flow control of a separated flow over a NACA 0015 airfoil and found that the actuation is able to delay stall and improve post-stall aerodynamic performance.  With plasma actuators, these control effects are also observed by \citet{PostCorke:AIAAJ2004} on the same airfoil. The uses of plasma actuators on other shear layer setups have been continuing to attract research interests \citep{Samimy:JFM2007,Corke:JA09,AkinsLittle:AIAA2015}.  Synthetic-jet-based flow control of a shear layer that forms in a separated flow over a wall-mounted hump is studied by \citet{Greenblatt:AIAAJ2006_P2}, who showed control input from the jet is able to change the generation rate of turbulent kinetic energy along the shear layer.  For shear layers in cavity flows, \citet{Cattafesta:PAS08} summarized flow control attempts using a number of modern actuators.  Flow over a backward-facing step is also a canonical problem setup where a shear layer is formed, and was chosen by \citet{VukasinovicGlazer:JFM2010} to investigate high-frequency actuation effects from a spanwise arrangement of synthetic jets.  While the above studies do not encompass all studies to date, they show the breadth of active control techniques being applied to shear layers that arise from a range of flows.

Amongst the aforementioned flows, the free shear layer formed by the mixing of two streams with different velocities is considered especially to be the canonical configuration for shear layer flows.  Studies on the characteristics of free shear layer dates back to the seminal work of \citet{BrownRoshko:JFM1974}, where they showed that the large spanwise roll-up coherent structures forming behind the trailing edge of a splitter plate serve as the main driving force for mixing of two streams and induce the entrainment that feeds fluid from two streams to the mixing region.  Along with their work, \citet{Winant74} and \citet{HoHuang:JFM82} have attributed the initial shear layer roll-up to the fundamental instability waves, where as their subharmonic instability accounts for the vortex merging downstream.  These findings are extended to compressible flow later by \citet{ElliottSamimy:PoF1990} and \citet{ClemensMungal:JFM1995}.  

As fundamental studies on the free shear layer instabilities, \citet{HoHuang:JFM82} introduced (non-local) subharmonic velocity disturbances in the entire free streams. \citet{Bechert:JFM1988_B} found the external acoustic excitation is also able to trigger the instability wave in the shear layer.  While leveraging flow instabilities is a general strategy for flow control, introducing control inputs in a limited local region may be a more practical approach than introducing control inputs externally.  On the other hand, \citet{BaroneLele:JFM2005} studied the receptivity of the shear layer behind a finite-thickness splitter plate to various type of disturbances by performing adjoint simulations.  Temperature disturbance is among one of them and the shear layer is found to be receptive to such perturbation.  However, studies on the excitation of shear layer instability by directly manipulating the local temperature are limited in the literature.  Based on these two reasons, the use of a localized thermal forcing input for flow control on a free shear layer draws our interests to the present study. 

The objective of this study is to assess the effectiveness of the use of local periodic heating for modifying the early evolution of a free shear layer.  Two-dimensional Navier--Stokes equations are numerically solved to simulate a spatially developing shear layer downstream a finite-thickness splitter plate, with local periodic heating introduced at the trailing edge.  As it has been pointed out by \citet{Crighton:ARFM1985} and \citet{Bechert:JFM1988_A}, active control of shear layer by introducing control inputs downstream of the trailing edge appears ineffective, but control via manipulation of the instabilities at the onset of a shear layer can be effective.  A slight perturbation during the genesis of a shear layer can lead to the overall change in flow physics downstream, because its instability characteristics determine which existing disturbances can be amplified.  Nonlinear effect takes place when disturbance becomes large in magnitude and transfers energy among modes while creating new ones, which drives the evolution of the flow.  Thus, focusing forcing input to the origin of a shear layer is favorable for flow control since the input can leverage the flow instabilities to grow and alter the flow behavior downstream.   

Our fundamental study on the use of local periodic heating as a forcing input to modify the spatial evolution of a shear layer is inspired by the development in thermophones and nanosecond pulse driven dielectric barrier discharge (ns-DBD) actuators.  For the former, thermophone is a sound generation technique discovered by \citet{Arnold:PRB17}.  With the application of alternating current (AC) over a platinum film, they found that the Joule heating on the film periodically increases the surrounding fluid temperature and leads to local pressure fluctuations at the frequency of alternating current.  This thermophone technique was not widely utilized in the past due to the relatively large power required for a desired surface pressure disturbance level.  In recent years, however, this drawback has been overcome by the use of graphene- and carbon-nanotube-based membranes for thermophones.  Compared to traditional metal membranes, the carbon-based membranes can be fabricated extremely thin, usually around $O(10^{-5})$ mm \citep{Tian:ACSNANO11}, such that its heat capacity per unit area (HCPUA) is at least two orders of magnitude lower than that of the metal films.  Since the output power of pressure disturbance delivered to the surrounding flow field is inversely proportional to the material HCPUA, the use of carbon-based membranes dramatically enhances the device efficiency.  The operating frequency is also extended in range ($20$ to $50$ kHz) with these membranes \citep{Tian:ACSNANO11,Bin:JAP2015}.  Considering thermophone as a new candidate for flow control actuators, the broadened range of its operating frequency is able to encompass that of many flow control applications.  

On the other hand, the use of ns-DBD actuators for flow control has also become widespread in recent years \citep{Little&Samimy:AIAAJ2012, Lehmann&Little:AIAA2014}.  The control mechanism of ns-DBD plasma actuation is believed to primarily rely on the deposition of thermal energy \citep{Nudnova:PPR2010,Adamovich:AIAA2012}, in contrary to alternating current driven dielectric barrier discharge (ac-DBD) actuators that employ electrohydrodynamic effect to introduce momentum perturbation \citep{Abe:AIAAJ2008,Corke:ARFM10}.  Numerous studies have reported the rapid heating effect near the pulse-driven plasma actuator and attributed the kinetic mechanism of the localized heating to the energy transfer to molecular translational/rotational modes on a sub-microsecond time scale \citep{Aleksandrov:JPD2010,Popov:JPD2011}.  For both thermophone and ns-DBD actuators, their energy-flux-based actuators rely on no mechanical moving parts. Moreover, their sheet-like arrangement facilitates surface-compliant installation without occupying any internal space or adding significant weight.  This advantage in installation allows them to be mounted on virtually any surface, and extend their potential uses for flow control on applications including rotorcraft blades and small-scale unmanned air vehicle wings.

In this study, we numerically examine the capability of this thermal energy-flux-based actuation for flow control on a canonical free shear layer flow.  In what follows, we describe the problem and discuss the computational setup in section \ref{sec:Comp_setup}.  Two boundary conditions for modeling local periodic heating are presented in section \ref{sec:HeatFluxBC}.  Two-dimensional baseline shear layers are characterized in section \ref{sec:BaselineChar}.  In section \ref{sec:CtrlMech}, the underlying mechanism of how thermal actuation can introduce perturbations to the flow field and alter the characteristics of the downstream shear layer is revealed by examining the near-field effect of the actuator.  The flow control effect by local periodic heating on the downstream shear layer is discussed in detail in section \ref{sec:control_effect}.  We conclude this study in section \ref{sec:conclusion} by offering summarizing remarks.

\section{Computational setup}
\label{sec:Comp_setup}

\subsection{Problem description}
\label{sec:ModelProblem}
We consider a two-dimensional free shear layer as a model problem to examine the effectiveness of thermal actuation.  Two incoming streams, initially separated by the splitter plate, at Mach numbers $M_1 = U_1/a_\infty = 0.4$ (top) and $M_2 = U_2/a_\infty = 0.1$ (bottom) with the same free stream acoustic speed, $a_\infty$, and pressure, $p_\infty$, form a spatially developing shear layer from the trailing edge, as illustrated in figure \ref{fig:setup}.  Following the definition by \citet{PapamoschouRoshko:JFM1988}, the mean convective Mach number for the shear layer is $M_c = \left( U_1 - U_2 \right)/\left( 2a_\infty \right) = 0.15$, with the density ratio $\rho_1/\rho_2$ being unity ($\rho_1 = \rho_2 = \rho_\infty$).  The plate-thickness-based Reynolds number, $Re_w \equiv \rho_\infty \bar{U} w/ \mu_\infty$, is set to $2500$, where $\bar{U} = (U_1 + U_2)/2$ is the convection velocity, $w$ is the plate thickness and $\mu_\infty$ is the free-stream dynamic viscosity.  The boundary layers on both sides of the plate are prescribed with the same momentum thickness, $\theta_0$, which is chosen to be $5\%$, $10\%$ and $25\%$ of the splitter plate thickness, $w$, resulting in the momentum-thickness-based Reynolds number $Re_{\theta_0} \equiv \rho_\infty \bar{U} \theta_0/\mu_\infty$ to range from $125$ to $625$.


\subsection{Computational approach}
\label{sec:gov.eqs}
The governing equations for the free shear layer are the compressible Navier--Stokes equations: 
\begin{align}
\label{eq:NS}
	\frac{\partial \tilde{\rho}}{\partial \tilde{t}} + \frac{\partial}{\partial \tilde{x}_j} \left( \tilde{\rho} \tilde{u}_j\right) &= 0, \\
\label{eq:NS2}
	\frac{\partial}{\partial \tilde{t}} \left( \tilde{\rho} \tilde{u}_i\right) + \frac{\partial}{\partial \tilde{x}_j} \left( \tilde{\rho} \tilde{u}_i \tilde{u}_j\right) &= - \frac{\partial}{\partial \tilde{x}_j} \left( \tilde{p} \delta_{ij} \right) + \frac{1}{Re} \frac{\partial}{\partial \tilde{x}_j}\left( \frac{\partial \tilde{u}_i}{\partial \tilde{x}_j} + \frac{\partial \tilde{u}_j}{\partial \tilde{x_i}} - \frac{2}{3}\frac{\partial \tilde{u}_k}{\partial \tilde{x_k}}\delta_{ij} \right), \\
\label{eq:NS3}
	\frac{\partial \tilde{e}}{\partial \tilde{t}} + \frac{\partial}{\partial \tilde{x}_j} \left[ \left( \tilde{e} + \tilde{p} \right) \tilde{u}_j\right] &=  \frac{1}{Re} \frac{\partial}{\partial \tilde{x}_j}\left[ \tilde{u}_i \left( \frac{\partial \tilde{u}_i}{\partial \tilde{x}_j} + \frac{\partial \tilde{u}_j}{\partial \tilde{x_i}} - \frac{2}{3}\frac{\partial \tilde{u}_k}{\partial \tilde{x_k}}\delta_{ij} \right)\right] + \frac{1}{Re Pr} \frac{\partial^2 \tilde{T}}{\partial \tilde{x_k} \partial \tilde{x_k}},
\end{align}
with the equation of state for ideal gas:
\begin{equation}
\label{eq:EOS}
	\tilde{p} = \frac{\tilde{\rho}\tilde{T}}{\gamma}, 
\end{equation}
where $\gamma$ is the specific heat ratio.  The non-dimensional variables in equations \ref{eq:NS} to \ref{eq:EOS} are the spatial coordinate $\tilde{x}_i$, time $\tilde{t}$, density $\tilde{\rho}$, velocity $\tilde{u}_i$, pressure $\tilde{p}$, energy $\tilde{e}$ and temperature $\tilde{T}$ according to the non-dimensionalization of
\begin{equation*}
	\tilde{\rho} = \frac{\rho}{\rho_\infty},~~\tilde{p} = \frac{p}{\rho_\infty a_\infty^2},~~\tilde{T} = \frac{T}{T_\infty},~~\tilde{e} = \frac{e}{\rho_\infty a_\infty^2},
\end{equation*} 
\begin{equation*}
	\tilde{u}_i = \frac{u_i}{a_\infty},~~\tilde{x}_i = \frac{x_i}{w},~~\tilde{t} = \frac{t a_\infty}{w}. 
\end{equation*}
We note that the shear layer roll-up wavelength, $\lambda_n$, may also be used for the non-dimensionalization of the spatial coordinate $x_i$ in order to provide fair comparisons between different $\theta_0$.  In this paper, it will be clearly stated which length scale is being used for non-dimensionalization.  The dimensionless parameters appearing in the governing equations are the acoustic Reynolds number and Prandtl number, given by
\begin{equation*}
	Re \equiv \frac{\rho_\infty a_\infty w}{\mu_\infty}~~\text{and}~~Pr \equiv \frac{\mu_\infty}{\rho_\infty\alpha_\infty}, 
\end{equation*}
where $\alpha_\infty$ is the free stream thermal diffusivity. 

To simulate the flow, we consider the two-dimensional set of equations and use the compressible flow solver CharLES \citep{Khalighi:ASME2011,Khalighi:AIAA11,Bres:AIAAJ2017}, which utilizes a second-order-accurate finite-volume method and a third-order Runge--Kutta scheme for time stepping. We use $\gamma = 1.4$, $Re = 10,000$ and $Pr = 0.7$ in the present study. Both values are representative of standard air. The temperature-varying dynamic viscosity, $\mu(T)$, is evaluated by the power law as $\mu = \mu_\infty ( T/T_\infty )^{0.76}$ \citep[see][]{Garnier:2009LES}.  The power law models the dynamic viscosity variation well for standard air in the range of $T/T_\infty \in [0.5, 1.7]$.  This range is suitable for the current study with local thermal inputs, where we observe the temperature fluctuation is within $22\%$ about $T_\infty$.

\begin{figure}
	\includegraphics[width=1.0\textwidth]{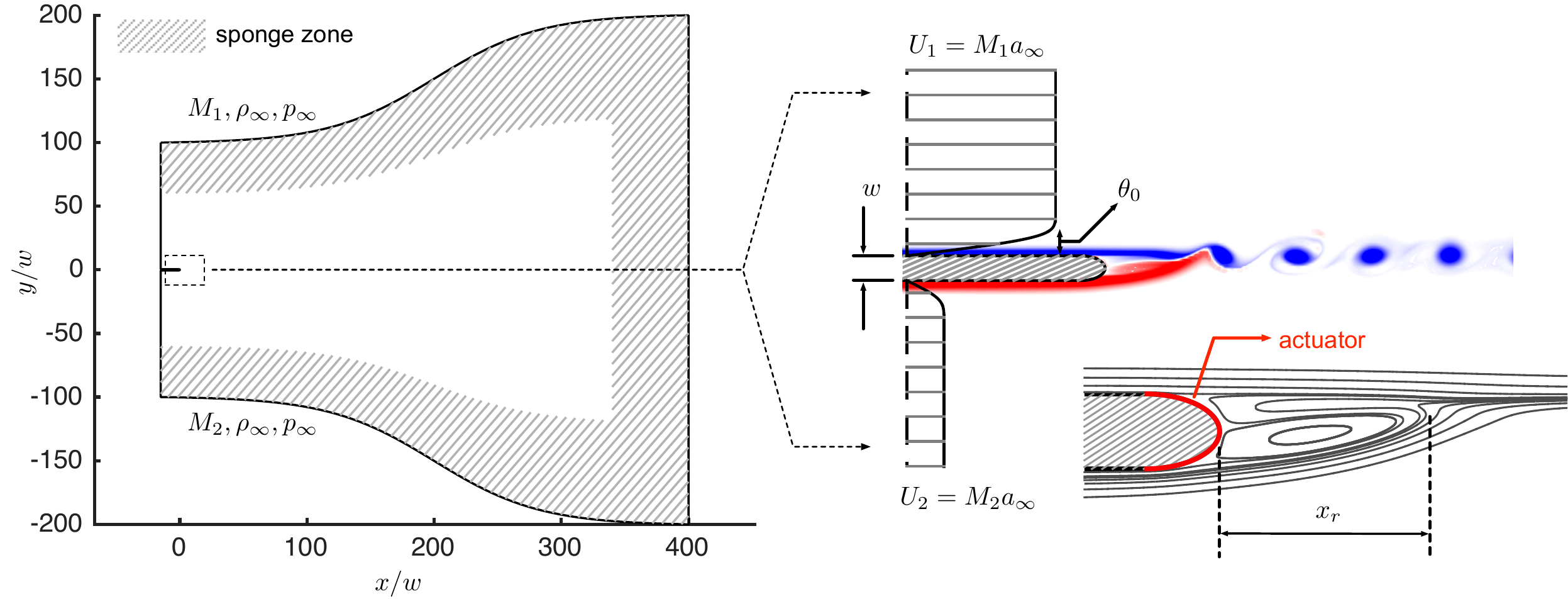}
	\caption{\label{fig:setup} Computational setup with spatially developing shear layer downstream of a splitter plate. Two streams at free-stream Mach number $M_1 = 0.4$ (top stream) and $M_2 = 0.1$ (bottom) enter the computational domain from the inlet boundary ($x/w = -15$) on the left. Sponge zone is provided in the shaded region of the computational domain. No-slip adiabatic boundary condition is imposed over the surface of the splitter plate. For forced flow, periodic heating is introduced from the actuator placed at the elliptic trailing edge by specifying an oscillatory heat flux boundary condition at a frequency $f^+$.}
\end{figure}

The computational domain, as illustrated in figure \ref{fig:setup} (left), has a streamwise extent of $x/w \in [-15, 400]$ and a maximum vertical extent of $y/w \in [-200, 200]$, similar to the domain setup used by \citet{SharmaLele:AIAA2011}. The splitter plate has an elliptic trailing edge with eccentricity of $0.866$, with its tip positioned at the origin, $x/w = y/w = 0$. The domain is discretized with a structured body-fitted mesh around the splitter plate.  The near-wall resolution is $\Delta y^+ \equiv \Delta y u^*/\nu_\infty = 0.25$ \citep[see][]{ChoiMoin:POF2012}, where $u^* = \sqrt{\tau_0/\rho_\infty}$ is the wall shear velocity and $\tau_0$ is the wall shear stress. To resolve the interaction between the shear layer and the localized periodic heat forcing, the wall-normal grid stretching rate is limited such that the $\Delta y^+$ only reaches a maximum of $1.25$ within $50w$ from the wall.  The overall grid size is approximately $4.4 \times 10^5$.  The DNS results from this mesh were verified by comparing the flow field to that from a finer mesh, where the grid is refined in the sheared region and has the size of $6.8\times 10^5$ grid points. No noticeable differences are found in the results from these two meshes.

Two streams are introduced to the computational domain from the inlet boundary (left) using Blasius boundary layer velocity profile.  Their density profiles are computed using the Crocco--Busemann relation, with the same far-field temperature and pressure.  Sponge zone is applied along the top, bottom and outlet (right) boundaries to damp out existing acoustic waves and vortical structures \citep{Freund:AIAAJ97}.  The target state of the sponge zone at the top and bottom boundaries are set to $[U_1, \rho_\infty, p_\infty]$ and $[U_2, \rho_\infty, p_\infty]$, respectively.  At the outlet, the time-average flow is set to be the target state of the sponge zone.

In this study, the effects of momentum thickness, the forcing frequency, and amplitude are examined in the context of modifying the characteristics of the shear layer. The flow control setup using local periodic heating and the associated input parameters are discussed below.

\subsection{Local periodic heating}
\label{sec:HeatFluxBC}
The thermal actuator model is implemented in the numerical simulations through a boundary condition.  Over the elliptic trailing edge of the splitter plate, an oscillatory energy-flux (unsteady Neumann) boundary condition that introduces local periodic heating is prescribed to the energy equation \ref{eq:NS3}, along with no-slip boundary condition for the momentum equation \ref{eq:NS2}.  In the present study, we consider two forms of periodic heating.  First, we consider a temporally oscillatory heat flux with zero mean as:

\begin{equation}
\label{eq:ZeroMeanForcing}
  \dot{q} = \hat{q} \sin( 2 \pi f^+ t) \cos \left( \frac{\pi}{w} y \right),~~~\left| y \right| < \frac{1}{2} w,
\end{equation}
where the net heating input to the local fluid is zero over the forcing period.  We also consider another forcing input $\dot{q}^p$ which introduces a positive direct current (DC) offset to the expression of $\dot{q}$ in equation (\ref{eq:ZeroMeanForcing}) such that the heat flux is always positive over the duty cycle:

\begin{equation}
\label{eq:PositiveForcing}
	\dot{q}^p = \hat{q} \left[ 1 + \sin( 2 \pi f^+ t) \right] \cos \left( \frac{\pi}{w} y \right),~~~\left| y \right| < \frac{1}{2} w.
\end{equation}
In the expressions for $\dot{q}$ and $\dot{q}^p$, $f^+$ and $\hat{q}$ denote the forcing frequency and amplitude, respectively.  For each choice of $\theta_0/w$, the forcing frequency, $f^+$, is normalized by the corresponding baseline shear layer roll-up frequency, $f_n$, and is varied from $f^+/f_n = 0.500$ and $1.25$, over which we observe forcing effects.  Since the heat flux amplitude, $\hat{q}$, accounts for the forcing power that is introduced to the shear layer flow, we normalize the total heating power from the trailing edge by the characteristic kinetic energy flux across the momentum thickness as 

\begin{equation}
\label{eq:NormalizedForcingPower}
	E^+ = \frac{Q}{\frac{1}{2} \rho_\infty \bar{U}^2 \cdot \theta_0 \bar{U}},
\end{equation}
where 

\begin{equation}
\label{eq:AvgForcingPower}
	Q = f^+ \int_0^{1/f^+} \int_{|y| < \frac{w}{2}} |\dot{q}| {\rm d}y {\rm d}t
\end{equation}
is the cyclic average heating power introduced by the actuator.  This normalization can account for the use of higher forcing amplitude, $\hat{q}$, for thicker initial momentum thickness, $\theta_0$, for a fair comparison of forcing requirements across different $\theta_0$, since thicker shear layer is expected to necessitate higher level of perturbation inputs to modify the shear layer.  We parameterize the normalized forcing power in this study in the range of $E^+ = 0.296$ to $2.96$.  This value is of the same order of magnitude of those used in plasma-based flow control works \citep{Samimy:JFM2007,Corke:ARFM10,SinhaSamimy:PoF2012,AkinsLittle:AIAA2015}.  We also note that the maximum temperature fluctuation observed over the surface of the actuator is within $22\%$ about the free stream temperature with the use of the largest forcing amplitude, $E^+ = 2.96$, in $\theta_0/w = 0.25$ cases.

\section{Baseline characterization}
\label{sec:BaselineChar}
\subsection{Shear layer characteristics}
\begin{figure}
	\begin{center}
	\includegraphics[scale=0.535]{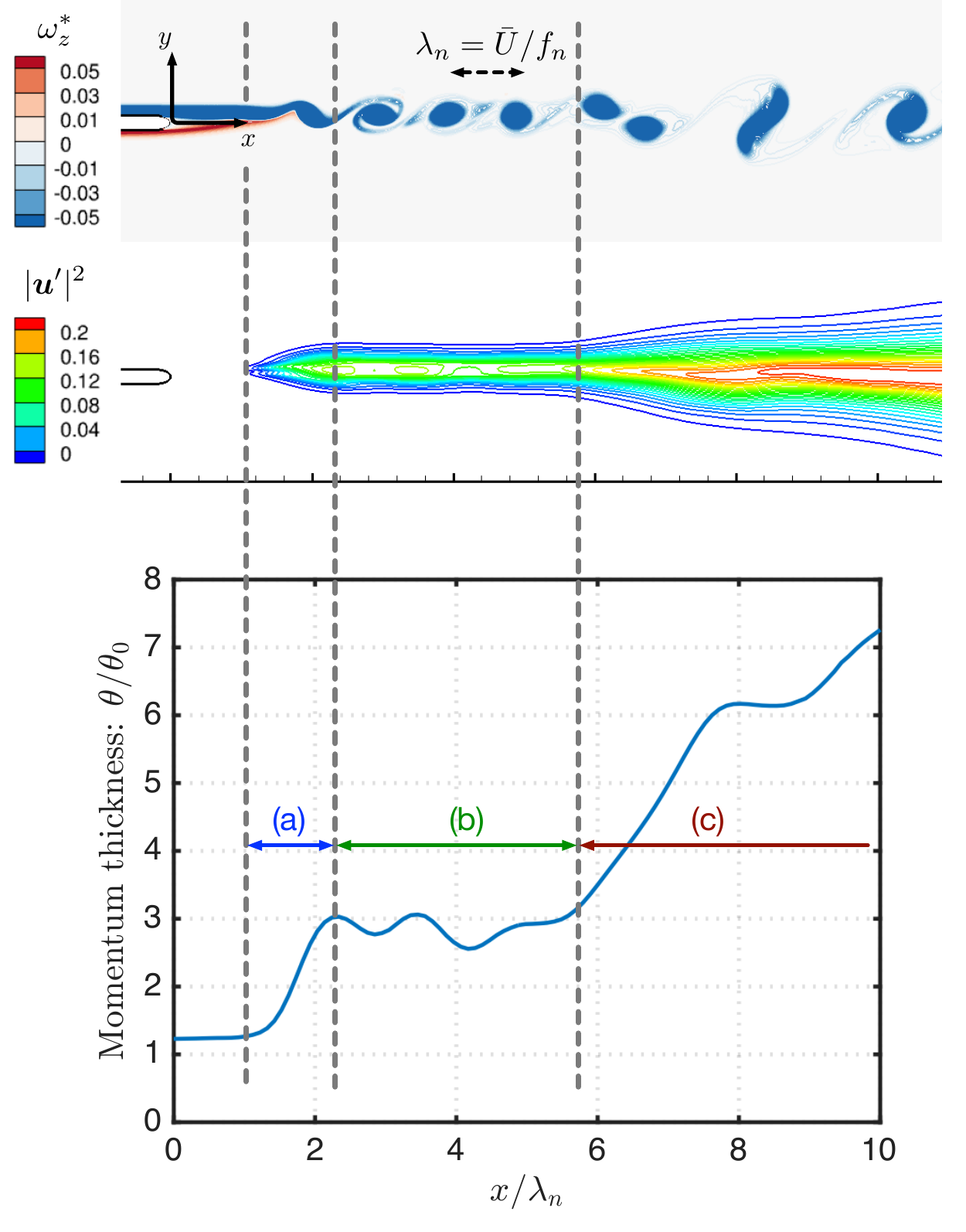}
	\end{center}
	\caption{\label{fig:BaselineChar}Baseline flow characterization for $\theta_0/w = 0.1$. Top: an instantaneous snapshot of normalized spanwise vorticity $\omega_z^* = \omega_z \theta_0/\bar{U}$; Middle: time-average velocity fluctuation magnitude $||\boldsymbol{u}'||^2 = (u_x'^2 + u_y'^2)/\bar{U}^2$; Bottom: spatial growth of momentum thickness ($\theta/\theta_0$). All share the same normalized streamwise coordinate, $x/\lambda_n$, where $\lambda_n = \bar{U}/f_n$ is the baseline roll-up wave length.}
\end{figure}

Let us first characterize the baseline flows at three momentum thicknesses specified at the inlet: namely, $\theta_0/w = 0.05$, $0.1$ and $0.25$. We focus our interest in the region within the streamwise extent of $10\lambda_n$, where $\lambda_n = \bar{U}/f_n$ is the fundamental shear layer roll-up wavelength. Shown in figure \ref{fig:BaselineChar} are a snapshot of the normalized instantaneous spanwise vorticity (top), the velocity fluctuation magnitude (middle), and the momentum thickness profile (bottom) for the baseline flow with $\theta_0/w = 0.1$. In this work, the momentum thickness is computed using the density and streamwise velocity profiles in the shear layer as
\begin{equation}
\label{eq:MomThickness}
	\theta(x) = \int_{-\infty}^{\infty} \frac{\overline{\rho}(x,y)}{\rho_\infty} \frac{\left[ \overline{u}(x,y) - U_2 \right]\left[ U_1 - \overline{u}(x,y) \right]}{\left( U_1 - U_2\right)^2} {\rm d}y,
\end{equation}
where the over-bar denotes the time-averaged quantity based on the flow statistics collected over more than 200 roll-up periods. Comparing the spatial growth of momentum thickness to the instantaneous flow field, we classify the shear layer flow into three regions: (a) the shear layer roll-up region, where the momentum thickness exhibits the first rapid growth behind the trailing edge; (b) the isolated vortex region, where each formed vortex remains compact as an individual and convects closely along the centerline; (c) the vortex merging region, where the vortices start to deviate from the centerline while convecting downstream, as the initiation of the vortex merging process. This last region can be identified by the second rapid growth of the momentum thickness. These three regions can be also characterized by the streamwise change in velocity fluctuation magnitude. In regions (a) and (c), the spatial extent and the magnitude of fluctuation are both increasing, but region (b) shows no apparent changes in the growth of the momentum thickness except for the rippling variation in the momentum thickness profile.  The cause of this ripple due to the orientation of the vortex evolving as the vortex convects downstream, which is discussed in appendix \ref{sec:mom.ripple}. Later in Section \ref{sec:control_effect}, the forcing effect of periodic heating on the characteristics of theses three regions are examined in further detail. 

\subsection{Influence of the splitter plate thickness}
Due to the finite thickness of the splitter plate, a small recirculation region of length $x_r$ develops directly behind the blunt trailing edge of the splitter plate, as illustrated by the time-average streamlines in figure \ref{fig:setup} (bottom-right).  This streamline pattern for two boundary layers merging behind the trailing edge of a finite-thickness splitter plate is also observed by \citet{SharmaLele:AIAA2011} and \citet{Laizet:PoF2010}.  The recirculation region can be viewed as a wake behind the finite-thickness splitter plate and result in a deficit in the profile streamwise velocity.  In the theoretical study by \citet{ZhuangDimotakis:PoF1995}, they reported that a wake instability mode can be introduced to the shear layer which has a deficit in its streamwise velocity profile due to a wake component, and the growth rates of both the wake mode and the fundamental shear layer mode increase when the streamwise velocity profile has a greater wake deficit. \citet{Mehta:EiF1991} experimentally showed that the splitter plate wake plays a dominant role in the development of the shear layer. The wake increases the turbulence levels in the near-field and shortens the development distance to achieve self-similarity in shear layer profiles.  Remarks by \citet{Mehta:EiF1991} on the lack of simple scaling for the shear layer with the presence of the wake also suggest that nonlinear interaction is taking place between the wake and the shear layer.  \citet{Laizet:PoF2010} performed DNS of shear layers developing behind three different trailing edge geometries and reported that the presence of the wake can lead to different turbulent structures and shear layer spreading rate.  We note, however, that there is no trace of von K{\'a}rm{\'a}n shedding observed in the simulations considered here, in spite of the presence of the recirculation region.

\begin{figure}
	\begin{center}
	\includegraphics[scale=0.535]{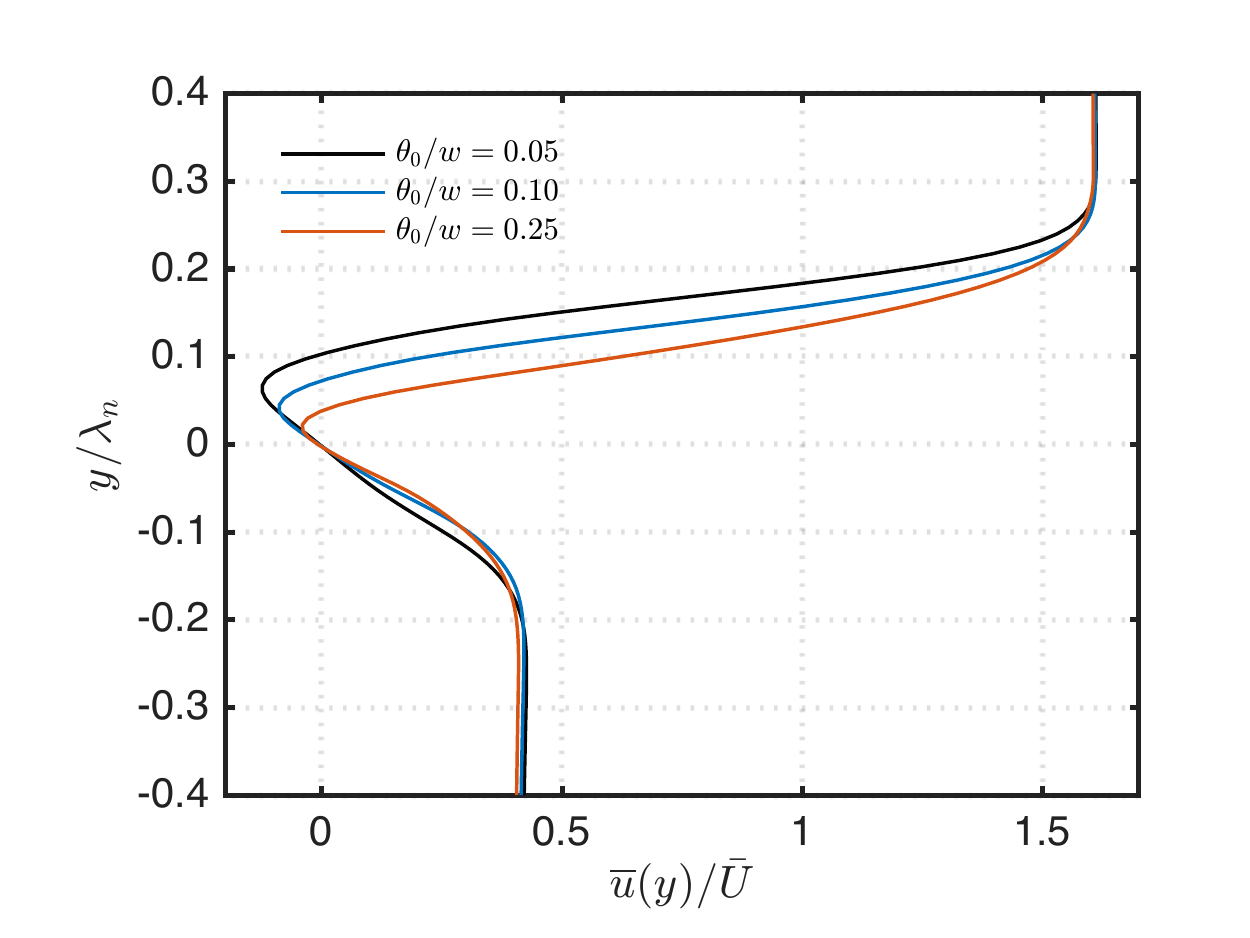}
	\end{center}
	\caption{\label{fig:Baseline_UAvg} Time-average streamwise velocity profiles at $x/w = 0.3$ over $y/\lambda_n$ for three baseline cases, showing stronger wake deficit for lower $\theta_0$.}
\end{figure}

\begin{table}
	\begin{center}
		\begin{tabular}{m{0.8in} m{0.5in} m{0.5in} m{0.5in} m{0.5in}}
			$\theta_0/w$	&$x_r/w$	&$f_n w/\bar{U}$	&$St_{\theta_0}$	&$\lambda_n/w$ \\
			\hline
			$0.05$	&$4.91$	&$0.282$	&$0.0141$	&$3.55$ \\
			$0.1$	&$4.84$	&$0.203$	&$0.0203$	&$4.92$ \\
			$0.25$	&$4.60$	&$0.112$	&$0.0279$	&$8.95$ \\
		\end{tabular}
	\end{center}
	\caption{\label{table:BaselineParas} Baseline characters of the shear layer for varied momentum thickness $\theta_0/w$: the length of recirculation region $x_r/w$; roll-up frequency $f_n w/\bar{U}$; roll-up Strouhal number $St_{\theta_0} = f_n \theta_0/\bar{U}$; and roll-up wavelength $\lambda_n/w$.}
\end{table}

\begin{figure}
	\begin{center}
		\begin{tabular}{c c}
			\includegraphics[scale=0.535]{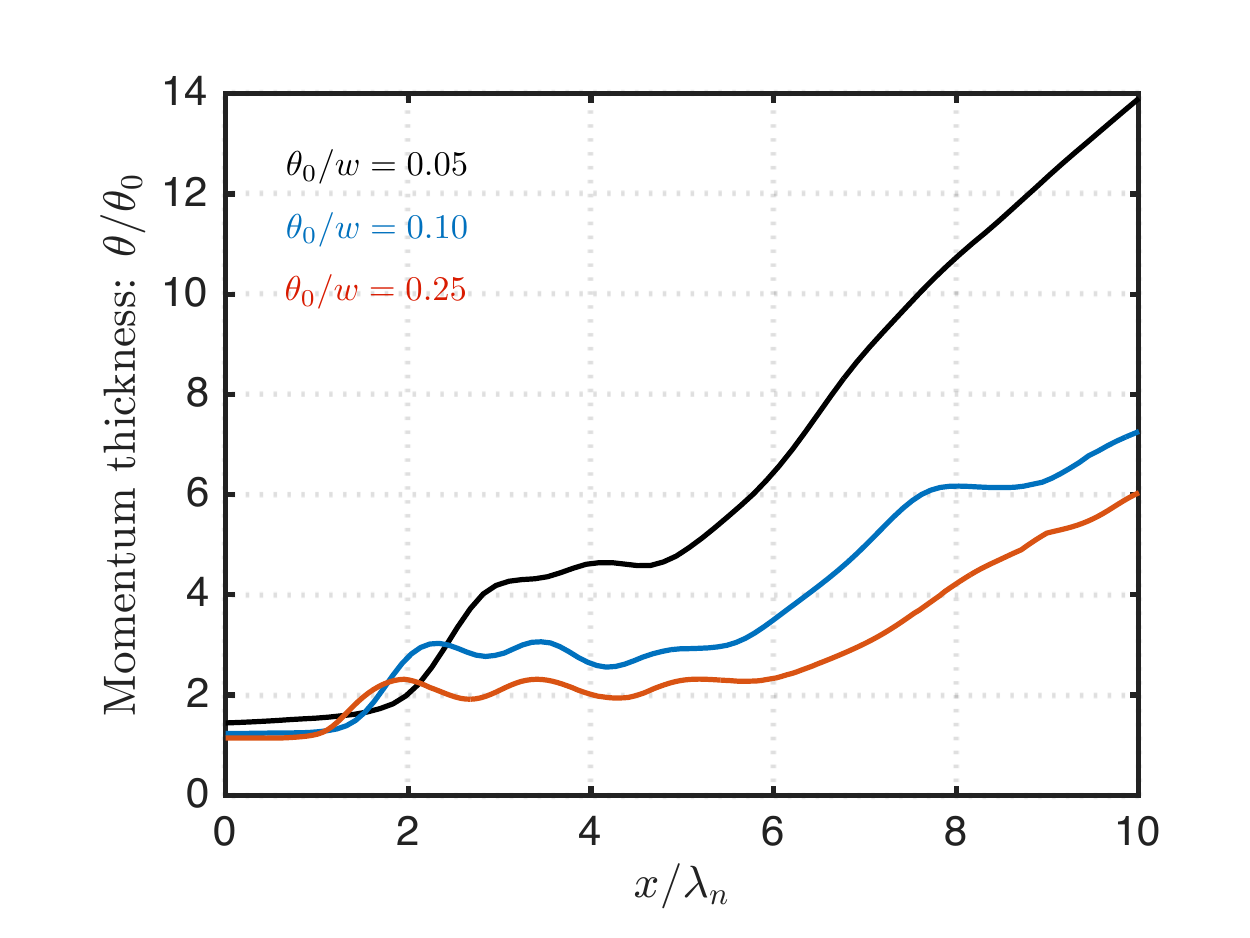} &
			\includegraphics[scale=0.535]{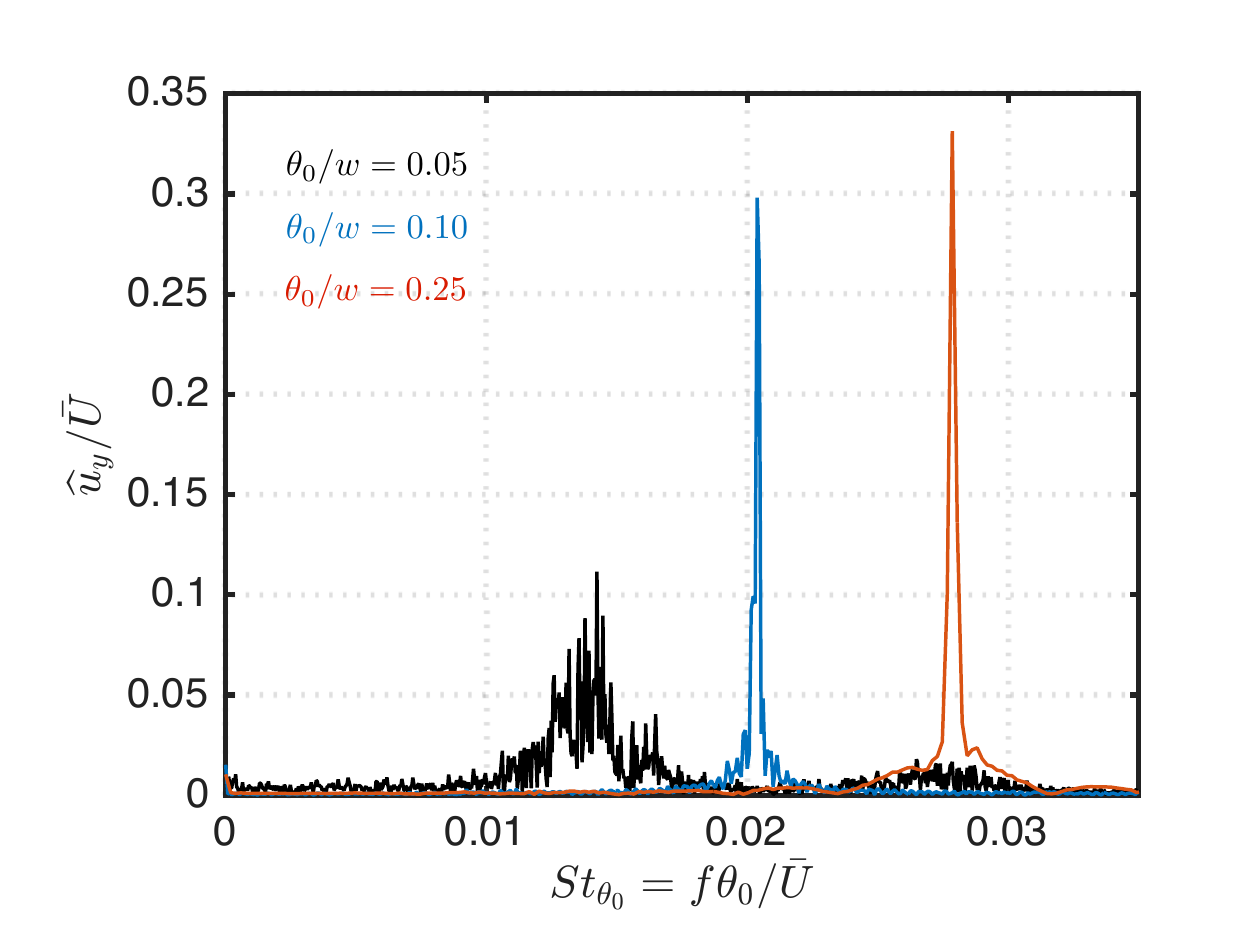}
		\end{tabular}
	\end{center}
	\caption{\label{fig:BaselineTheta&Specs}(Left) Momentum thickness growth for baseline flows with $\theta_0/w = 0.05$, $0.1$ and $0.25$. (Right) Transverse velocity spectra for three baseline cases at $x/\lambda_n = 2.5$. The shear layer roll-up frequency $f_n$ for each corresponding case is determined by the peak of each spectrum.}
\end{figure}

The lengths of the recirculation regions for each baseline cases with varied momentum thickness $\theta_0/w$ are summarized in Table \ref{table:BaselineParas}, where the change in the wake length, $x_r/w$, is limited to $6\%$ while $\theta_0/w$ is increased by five times. At the streamwise station of $x/w = 3.0$, the time-average streamwise velocity profile over the normalized transverse coordinate $y/\lambda_n$ is presented in figure \ref{fig:Baseline_UAvg}.  Note that the normalization of $y/\lambda_n$ causes the velocity minimum to appear at different location for different choice of $\theta_0/w$, since the value of $\lambda_n$ influenced by the value of $\theta_0/w$.  In figure \ref{fig:Baseline_UAvg}, we observe that the wake deficit increases in both its transverse extent and maximum velocity deficit for thinner incoming boundary layer.  In figure \ref{fig:BaselineTheta&Specs} we show the momentum thickness profile over the streamwise coordinate normalized by the corresponding roll-up wavelength for each case.  As discussed previously using $\theta_0/w = 0.1$ for the baseline characterization, the three-region growth patterns are also observed in the other two baseline cases. The initiation of merging takes place at $x/\lambda_n \approx 6$ for all cases. However, thicker $\theta$ for the isolated vortex region and delayed roll-up are observed for thinner $\theta_0/w$.  These observations can be attributed to the stronger effect of wake when $\theta_0$ is thinner.  The effect of the plate wake can be also seen in the shear layer roll-up Strouhal number, $St_{\theta_0} = f_n\theta_0/\bar{U}$, in Table \ref{table:BaselineParas}.  We also note that the values of $St_{\theta_0} $ are lower than the prediction from linear instability theory \citep{Monkewitz:PoF1982,HoHuerre:AR84}, $St_{\theta_0} = 0.032$, in all baseline cases.  This also suggests that the effective initial momentum thickness is thicker than that on the high speed side, $\theta_0$, because of the presence of the wake. Therefore, for thinner $\theta_0$ where the wake effect is more pronounced, the greater  deviation of $St_{\theta_0}$ from $0.032$ is observed.  The stronger wake effect on thin $\theta_0$ can also be inferred from figure \ref{fig:BaselineTheta&Specs}, where the transverse velocity spectra at $x/\lambda_n = 2.5$ for three baseline cases are shown.  The clean peaks in the cases of $\theta_0/w = 0.1$ and $0.25$ suggest a synchronized shear layer roll-up and shedding, whereas the cause of the less prominent peak and broader frequency content in the case of $\theta_0/w = 0.05$ can be attributed to the nonlinear interaction between the shear layer roll-up mode and the wake mode that redistribute the energy of the roll-up frequency to other nearby frequencies.  

\subsection{Local stability analysis}
\label{sec:InstabilityAnalysis}

To determine regions with linear growth of perturbation and predict the corresponding growth rate for the Kelvin--Helmholtz instability, local linear spatial stability analysis on the baseline flow is also performed at the onset region and the isolated vortex region of the shear layer.  We compute the corresponding spatial growth rate at a specified perturbation frequency using the time-average density and streamwise velocity profiles as the base flow for the compressible Rayleigh equation \citep{Sabatini:AIAAJ14}.  

\begin{figure}
	\begin{center}
	\vspace{0.2in}
		\includegraphics[scale=0.535]{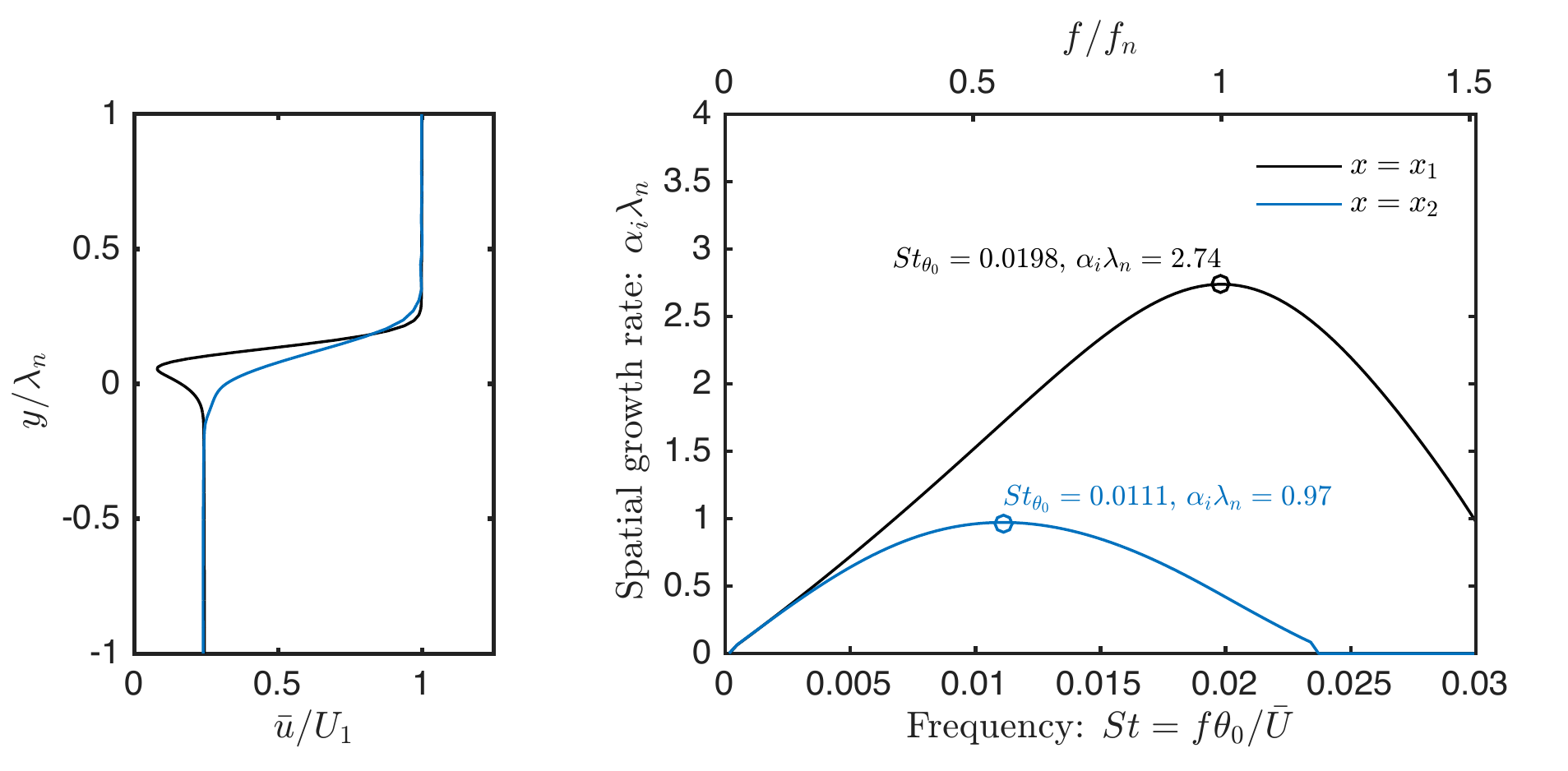}
	\end{center}
	\caption{\label{fig:BaselineGR}Spatial growth rates from linear stability analysis at two streamwise stations for $\theta_0/w = 0.1$ baseline. Both agree with results from DNS shown in figure \ref{fig:3DSpectrum}. Theoretical growth rates at forcing frequencies that appear in section \ref{sec:control_effect} are tabulated in table \ref{table:ForceModeGR}.}
\end{figure}

\begin{figure}
	\begin{center}
		\includegraphics[scale=0.55]{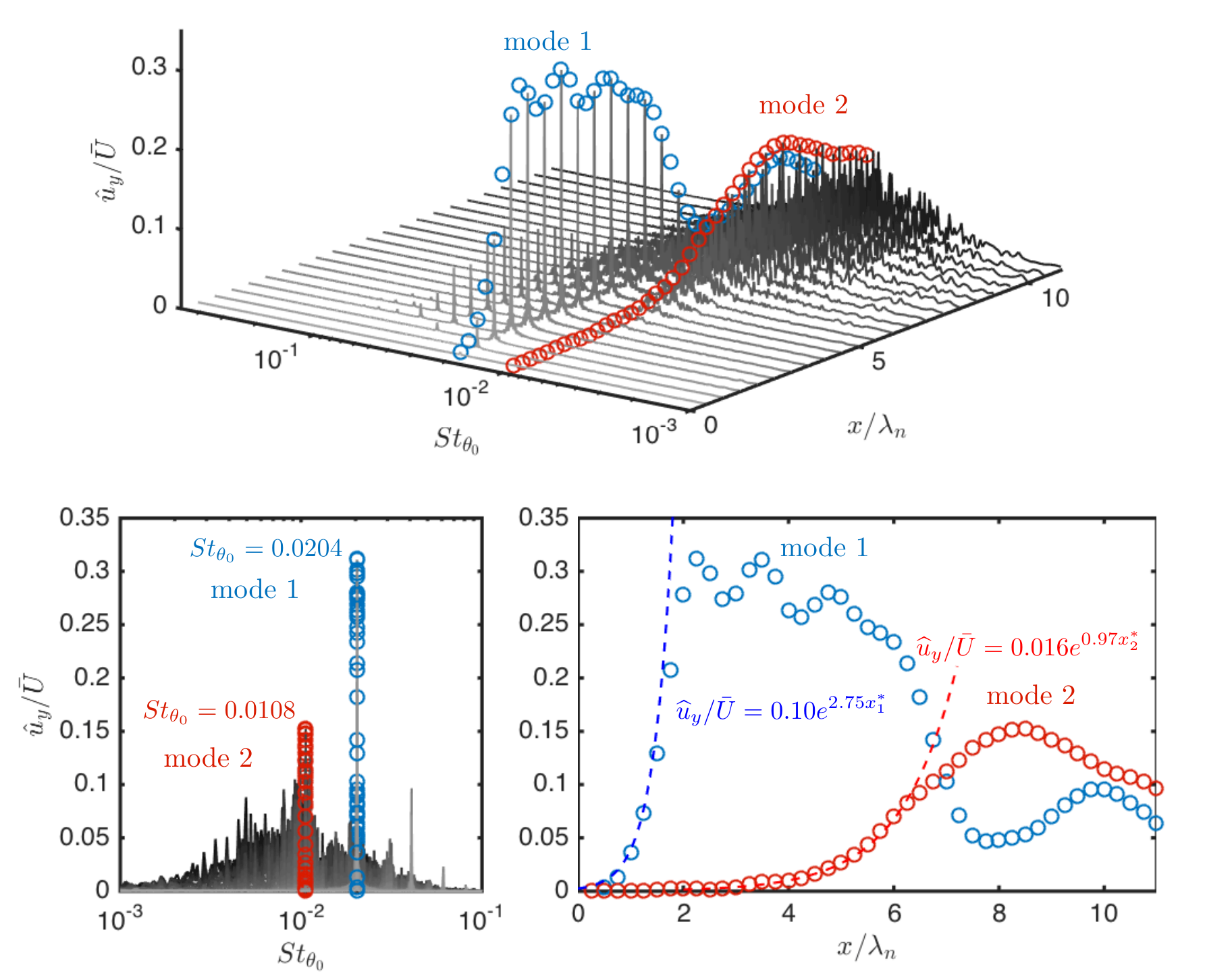} 
	\end{center}
	\caption{\label{fig:3DSpectrum}Transverse velocity spectra over the centerline: (top) the spatial development of spectra; (bottom-left) view from amplitude-frequency plane; (bottom-right) streamwise profiles of modal amplitudes for roll-up and merging mode with fitted exponential growth functions, where $x_1^* = (x-x_1)/\lambda_n$ and $x_2^* = (x-x_2)/\lambda_n$. The frequencies and growth rates of mode 1 and mode 2 agree with linear stability theory prediction in figures \ref{fig:BaselineGR} with less than $3\%$ differrence.}
\end{figure}

Figure \ref{fig:BaselineGR} shows the stability analysis results for $\theta_0/w = 0.1$ baseline flow.  Time-average flow profiles at two streamwise stations (shown on the left) are chosen about which to perform the instability analysis.  The first station $x_1$, corresponding to $\left( x_1 - x_r \right)/\lambda_n = 0.24$ or $x_1/\lambda_n = 1.22$, is located slightly downstream of the recirculation region but upstream of the shear layer roll-up region. In figure \ref{fig:BaselineChar}, the station $x_1/\lambda_n = 1.22$ is located in the flat region before the first sudden growth of the momentum thickness. Similarly, we choose the second station $x_2$, where $\left( x_2 - x_r \right)/\lambda_n = 3.1$ or $x_2/\lambda_n = 4.06$, in the isolated vortex region indicated by the second flat region and before the second sudden growth.  In figure \ref{fig:BaselineGR}, the most amplified frequencies and the corresponding spatial growth rates are $St_{\theta_0} = 0.0198, \alpha_i \lambda_n = 2.74$ and $St_{\theta_0} = 0.0111, \alpha_i \lambda_n = 0.97$ for stations $x = x_1$ and $x = x_2$, respectively.  We compare these values to the results from DNS shown in figure \ref{fig:3DSpectrum}. In the top sub-figure, the spectra of transverse velocity is plotted over a series of streamwise locations. The symbol {\color{blue2} $\boldsymbol{\circ}$} (referred to as mode 1 hereafter) represents the first spatially amplified mode and {\color{red2} $\boldsymbol{\circ}$} (referred to as mode 2 hereafter) depicts the second mode.  Their corresponding frequencies in terms of Strouhal numbers are $St_{\theta_0} = 0.0204$ and $0.0108$ for modes 1 and 2, respectively, as captured in the bottom-left figure. Compared to the frequencies predicted by linear stability theory, the differences are only approximately $1.5\%$ and $3\%$ for the modes 1 and 2, respectively.  The spatial growth of the modal amplitude can be seen in bottom-right subplot of figure \ref{fig:3DSpectrum}.  Also, in the bottom-right figure, the amplitude growth of the modes 1 and 2 are fitted with exponential growth function in $0 \le x/\lambda_n \le 1.6$ and $2.5 \le x/\lambda_n \le 5.5$, respectively for each mode. We also find the growth rates $\alpha_i \lambda_n$ from the fitting differ only by $0.7\%$ and $2\%$ from the prediction of the linear theory for roll-up and merging modes, respectively.  The fitted function deviates from the simulation data {\color{blue2}$\boldsymbol{\circ}$} and {\color{red2}$\boldsymbol{\circ}$} past $x/\lambda_n = 1.5$ and $6.5$, respectively.  These deviations are expected since, referring to figure \ref{fig:BaselineChar}, these two stations have already reached the vortex formation and merging regions where momentum thickness exhibits a sudden streamwise growth and the parallel flow assumption is no longer valid.

\begin{table}
	\vspace{0.2in}
	\begin{center}
		\begin{tabular}{m{0.4in} | m{0.3in} m{0.3in} m{0.3in} | m{0.3in} m{0.3in}}
			 						& \multicolumn{3}{c |}{$x = x_1$} & \multicolumn{2}{c}{$x = x_2$}\\
									\hline
			 \multicolumn{1}{c |}{$f^+/f_n$}				& $1.00$	& $1.16$	& $1.25$	& $0.500$	& $0.575$ \\
			 \multicolumn{1}{c |}{$\alpha_i\lambda_n$}		& $2.74$	& $2.50$	& $2.22$	& $0.931$	& $0.942$         
		\end{tabular}
	\end{center}
	\caption{\label{table:ForceModeGR} The theoretical spatial growth rates obtained at forcing frequencies appearing in section \ref{sec:control_effect} at stations $x = x_1$ and $x_2$. These values of growth rates will be compared to the growth rates of forcing mode later in figures \ref{fig:VortexTrackLow2} and \ref{fig:VortexTrackHigh2}.}
\end{table}

The agreement between the results from DNS and the linear stability analysis on the baseline flow provides deep insight into the growth of perturbations in the aforementioned regions.  As we discuss later, the frequencies in the vicinity of that of mode 2 are also considered as candidates for the actuation frequency of the local periodic forcing in Section \ref{sec:control_effect}.  The growth rate prediction at the forcing frequency will also be compared to the forced cases later in discussions related to figures \ref{fig:VortexTrackLow2} and \ref{fig:VortexTrackHigh2}.
 
\section{Thermal control mechanism}
\label{sec:CtrlMech}
In this section, the local effects of periodic thermal forcing is studied to understand how the thermal perturbations give rise to hydrodynamic perturbations and modifies the shear layer physics downstream.  The local temperature fluctuation and surface pressure disturbance introduced by local periodic heating change local vorticity flux through the solid boundary, either by changing the surrounding fluid properties or wall-tangential pressure gradient \citep{Hornung:Melbourn1989,WuWu:JFM1993}.  The fluctuations in pressure and density in the vicinity of the actuator may also generate volumetric baroclinic vorticity, $(\nabla \rho \times \nabla p) /\rho^2$.  In this two-dimensional study, we evaluate the local wall-normal vorticity flux using \citep{WuWu:JFM1993} 
\begin{equation}
\label{eq:VorticityFlux}
	\sigma_z(s) = - \frac{1}{\rho_0(s)} \left\lbrace \hat{\boldsymbol{e}}_n(s) \cdot \boldsymbol{\nabla} \left[ \mu\left(s\right) \omega_z\left(s\right) \right]_0 \right\rbrace,
\end{equation}
where $\mu$ and $\omega_z$ are the dynamic viscosity and spanwise vorticity, respectively, subscript $0$ denotes the wall adjacent quantities, $s$ is the spatial parameter describing the splitter plate surface, and $\hat{\boldsymbol{e}}_n$ is the unit wall-normal vector. In two-dimensional Cartesian coordinates, the baroclinic torque generation is computed using  
\begin{equation}
\label{eq:BaroGen}
	B_z(x,y) = \frac{1}{\rho^2} \left( \frac{\partial \rho}{\partial x}\frac{\partial p}{\partial y} - \frac{\partial \rho}{\partial y}\frac{\partial p}{\partial x} \right).
\end{equation}
These two quantities from the baseline and forced flows are examined below.

Without loss of generality, we consider the case with $\theta_0/w = 0.1$.  The two quantities in equations \ref{eq:VorticityFlux} and \ref{eq:BaroGen} for the baseline and forced flows are examined to quantify their fluctuation magnitudes and frequencies.  The forced and baseline computations are started at $tf_n = 0$ from the same initial condition, which is associated with the instantaneous baseline flow with steady flow statistics.  For the forced case, periodic heating at the trailing edge is turned on at $tf_n = 0$ using equation \ref{eq:ZeroMeanForcing} with $f^+/f_n = 0.500$ and amplitude of $E^+ = 0.741$.  Flow statistics are collected over $tf_n \in [100,250]$.

\begin{figure}
	\begin{center}
		\includegraphics[scale=0.535]{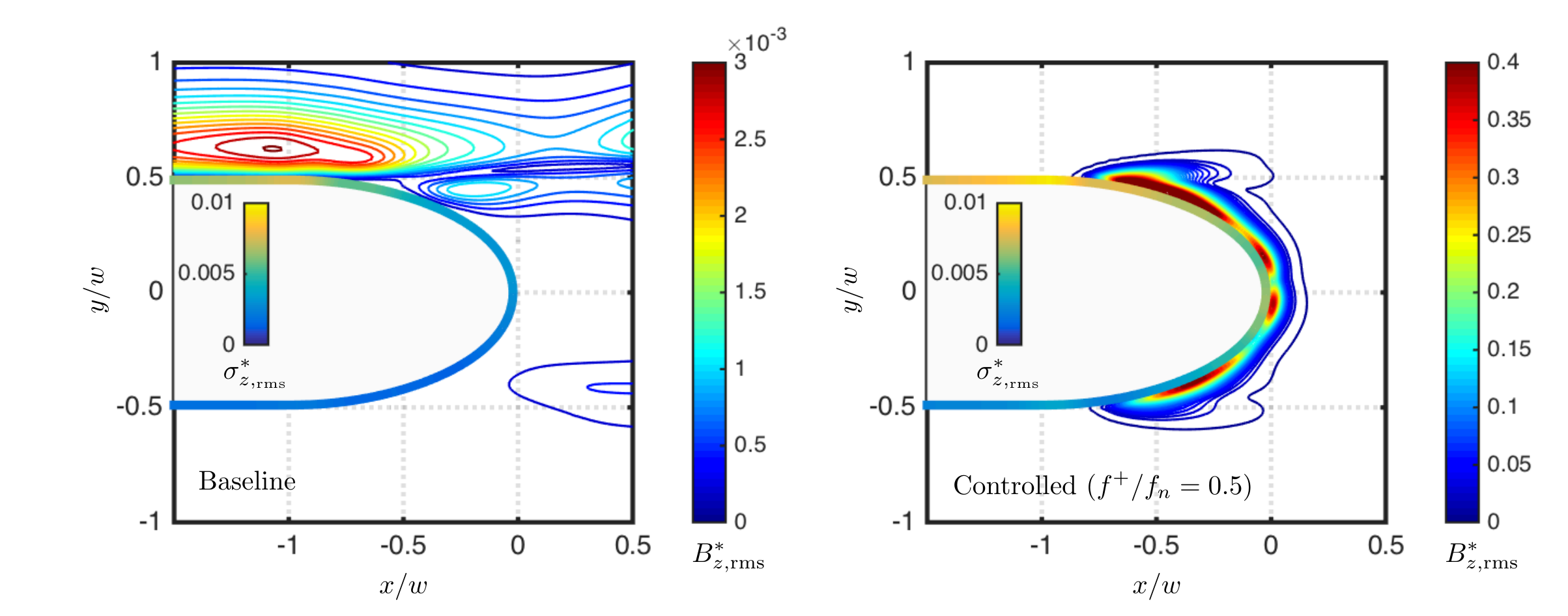}
	\end{center}
	\caption{\label{fig:ChangeInNearField} Fluctuation magnitudes of spanwise vorticity flux over the plate surface, $\sigma^*_{z,\text{rms}}$, and volumetric baroclinic torque generation rate, $B^*_{z,\text{rms}}$, of the baseline (left) and forced ($f^+/f_n = 0.500$, right) flows. Both quantities exhibit higher levels of fluctuations in the controlled case.}
\end{figure}

\begin{figure}
	\begin{tabular}{c c}
		\begin{overpic}[scale=0.535]{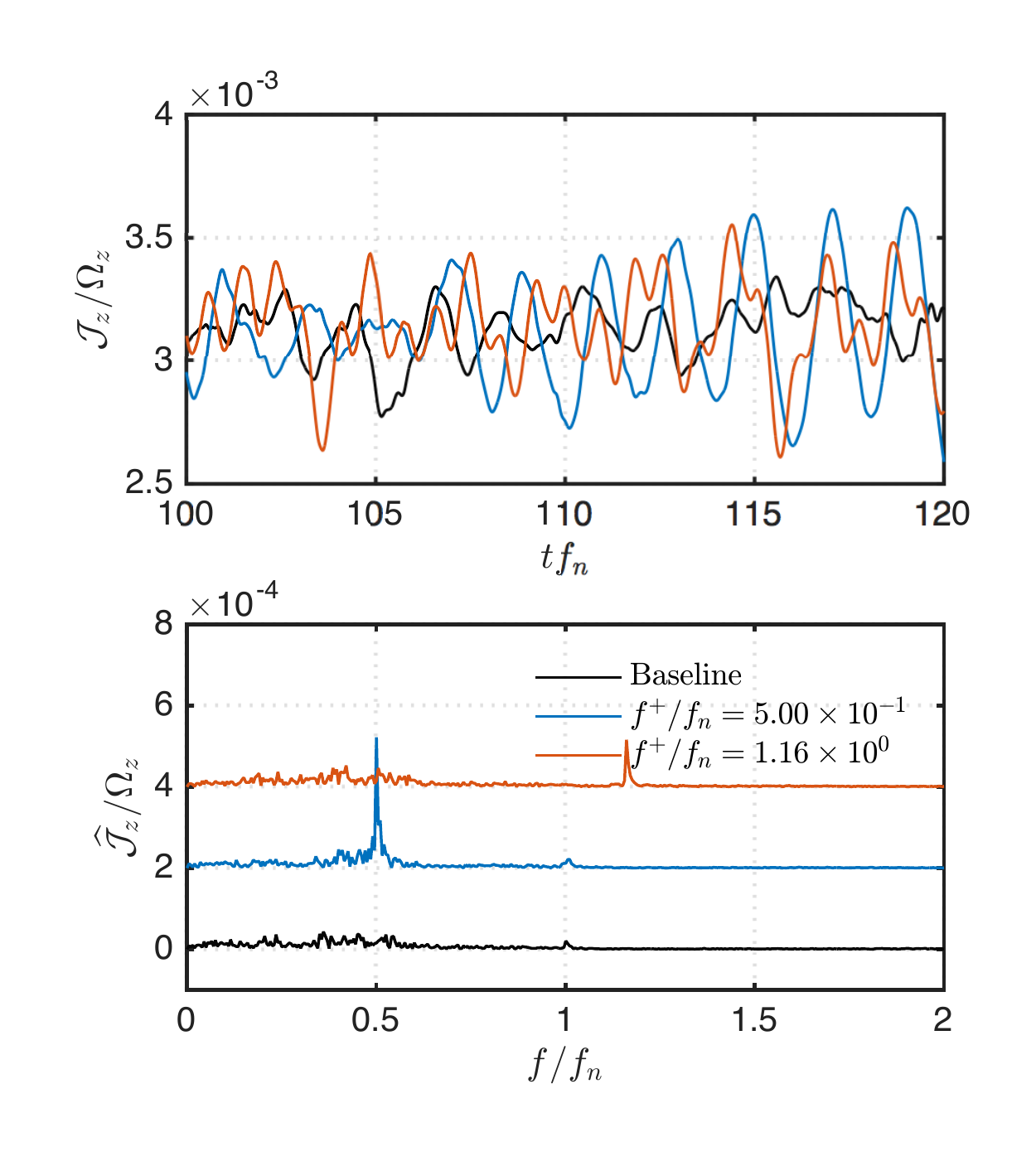}
		\end{overpic} &
		\begin{overpic}[scale=0.535]{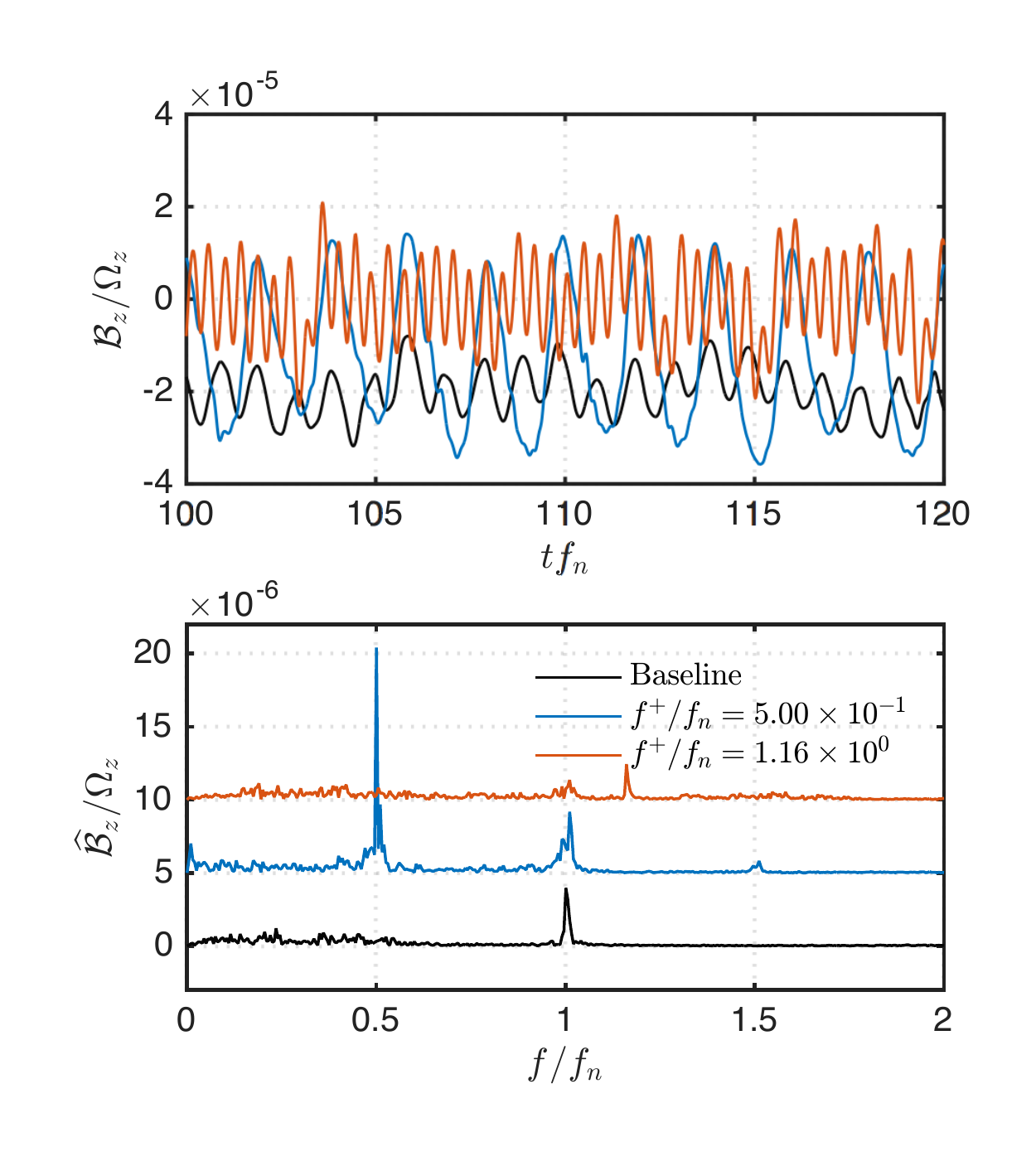}
		\end{overpic}
	\end{tabular}
	\caption{\label{fig:NoseFlux&BaroGen} Temporal oscillations (top) and their frequency spectra (bottom) for total vorticity flux $\mathcal{J}_z$ (left) over the trailing edge, and total volumetric baroclinic torque generation, $\mathcal{B}_z$ (right) near the trailing edge. Measurements from the baseline and two forced ($f^+/f_n = 0.500$ and $f^+/f_n = 1.16$) flows are presented. In the control cases, both $\widehat{J}_z/\Omega_z$ and $\widehat{B}_z/\Omega_z$ have a prominent peak at the forcing frequency. The spectra are shifted vertically for graphical clarity.}
\end{figure}
  
The change in these two quantities between the forced and baseline cases is shown in figure \ref{fig:ChangeInNearField}, where $\sigma^*_{z,\text{rms}} = \sigma_{z,\text{rms}}/\Omega_z$ is the root-mean-square value of $\sigma_z$, normalized by $\Omega_z = \int_{y \in \textbf{inlet}} \omega_z \boldsymbol{u} \cdot \hat{\boldsymbol{e}}_n {\rm d}y$, namely, the total Euler flux of vorticity introduced by two inlet boundary layers. Similar normalization is used for $B^*_{z,\text{rms}}$. We observe that both of these quantities exhibit higher fluctuation magnitudes in the thermally forced flow. Furthermore, note that the maximum magnitude of $B^*_{z,\text{rms}}$ in the vicinity of the actuator is $O(10^{2})$ compared to the baseline flow. As reported by \citet{Cheung:PoF09}, the baroclinic vorticity generation plays an important role in governing shear layer dynamics, especially for subsonic shear layers.  In their study, the order of magnitude difference between a heated (free-stream temperature ratio of $1.5$) and an unheated (with the same free-stream temperature) shear layer is also of $O(10^{2})$. The additional generation of baroclinic vorticity takes part in the vortex dynamics in the heated shear layer and leads to different behaviors from the unheated shear layer. 

We compute the total vorticity flux over the trailing edge as 
\begin{equation}
	\mathcal{J}_z = \int_{s \in \text{tip}} \sigma_z(s) {\rm d}s 
\end{equation}
and the total baroclinic generation near the trailing edge as 

\begin{equation}
	\mathcal{B}_z = \int_{|\boldsymbol{r}| < 2 w} B_z(\boldsymbol{r}) {\rm d}v. 
\end{equation}
The quantities $\mathcal{J}_z$ and $\mathcal{B}_z$ are normalized by $\Omega_z$ and plotted in figure \ref{fig:NoseFlux&BaroGen} over time and frequency domains.  As observed from the frequency domain, both of these quantities peak at the forcing frequencies, suggesting the fluctuations in $\mathcal{J}_z$ and $\mathcal{B}_z$ are indeed induced by the local periodic forcing.

The present flow control technique utilizes local periodic heating as an energy-deposition-based forcing, which introduces additional generation of vorticity.  As vorticity quantifies the rotation of fluid particles, we may also consider the present technique as a vorticity-based forcing, in contrast to other momentum-based flow control techniques such as synthetic jets \citep{Glezer:ARFM2002}. When the extra oscillatory vorticity is introduced to the shear layer at its onset, it can trigger the instability responsible for the shear layer roll-up, or modify the strength of each vortex formed from shear layer roll-up and change their interaction dynamics downstream.

\section{Flow control effects on the shear layer}
\label{sec:control_effect}
In this section, the influence of the forcing frequency ($f^+$) and the forcing type ($\dot{q}$ and $\dot{q}^p$) on the shear layer is examined.  Each DNS is initialized with an instantaneous flow field from the baseline flow, with steady statistics based on at least $300$ fundamental roll-up periods.  Forcing is turned on from this initial condition. Flow statistics for the controlled flows are collected after $100$ roll-up periods.  The reported controlled flow spectrum and mean-flow-based momentum thickness are evaluated over at least another $200$ roll-up periods.  We start the discussion with the use of $\dot{q}$ in section \ref{sec:LowFQForcing} and \ref{sec:HighFQForcing}. Forcing with $\dot{q}^p$, which is $\dot{q}$ with a positive direct current (DC) offset, is examined in section \ref{sec:PositiveMeanAct}.

\subsection{Spreading enhancement ($f^+ \approx 0.5f_n$)}
\label{sec:LowFQForcing}
We examine the forcing effect when the actuation frequency is close to the first subharmonic of the roll-up frequency with the use of $\dot{q}$. Two forcing frequencies are examined: namely, $f^+/f_n = 0.575$ and $0.500$. The forcing frequency $f^+/f_n = 0.575$ is selected for being close to the frequency of mode 2 found in figure \ref{fig:3DSpectrum}.  The exact first subharmonic $f^+/f_n = 0.500$ is also chosen.  For fair comparison, all cases presented in this section are based on the same forcing power of $E^+ = 0.741$.  In the discussions below, we first highlight the common forcing effects for these two forcing frequencies. We then discuss the differences between the two forced cases.

\begin{figure}
\vspace{1cm}
	\includegraphics[width=1.0\textwidth]{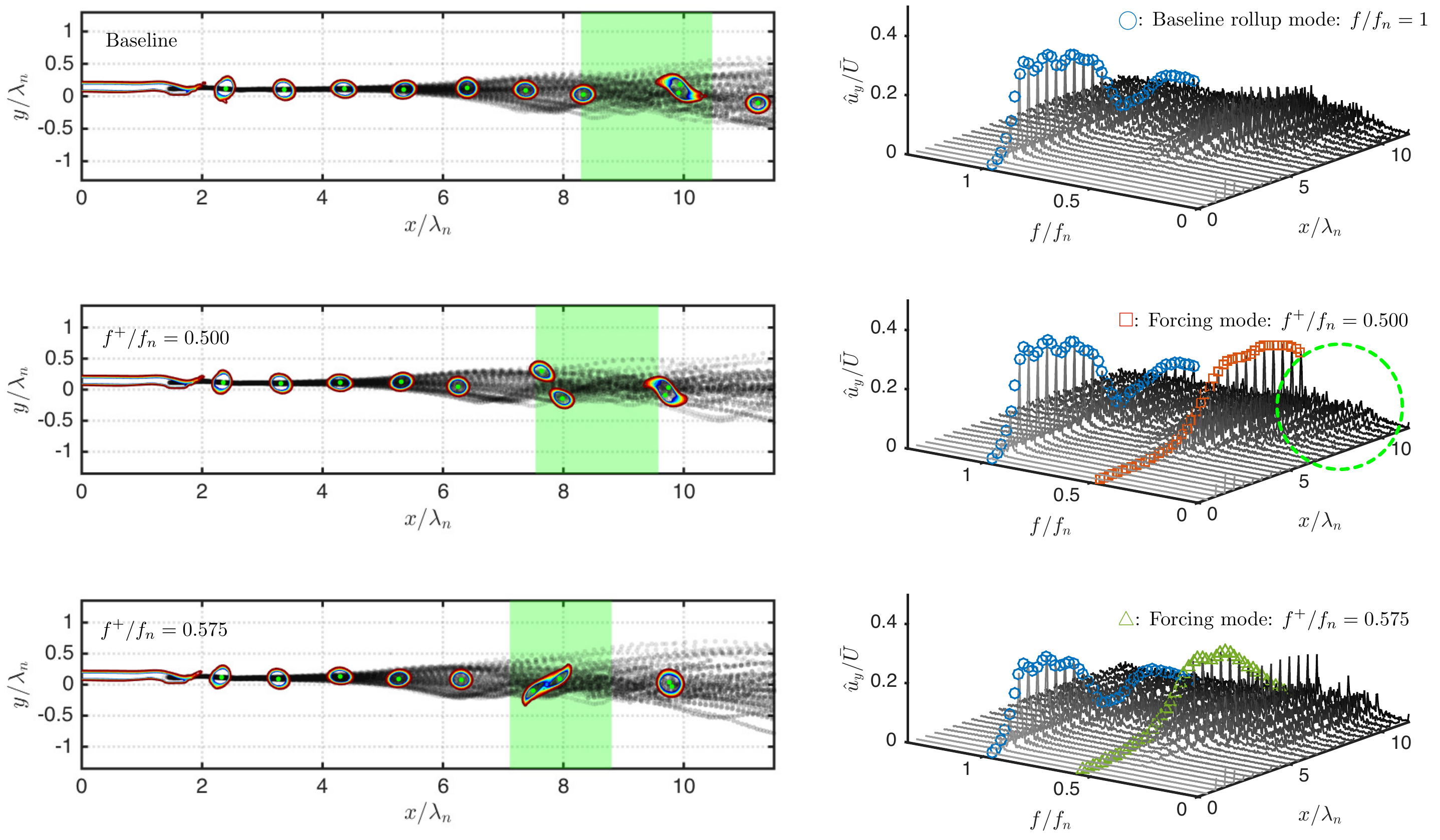}
	\caption{\label{fig:VortexTrackLow1} Vortex trajectories (left) and modal amplitude evolution (right) for $\theta_0/w = 0.1$. The merging location and its spatial variation is depicted by the green-shaded region (left) for each case. The forcing mode for $f^+/f_n = 0.500$ maintains its high level of modal amplitude while marching downstream, and suppresses other lower frequency modes to grow, as highlighted by the cleaner spectra in the green circled region.}
	\includegraphics[width=1.0\textwidth]{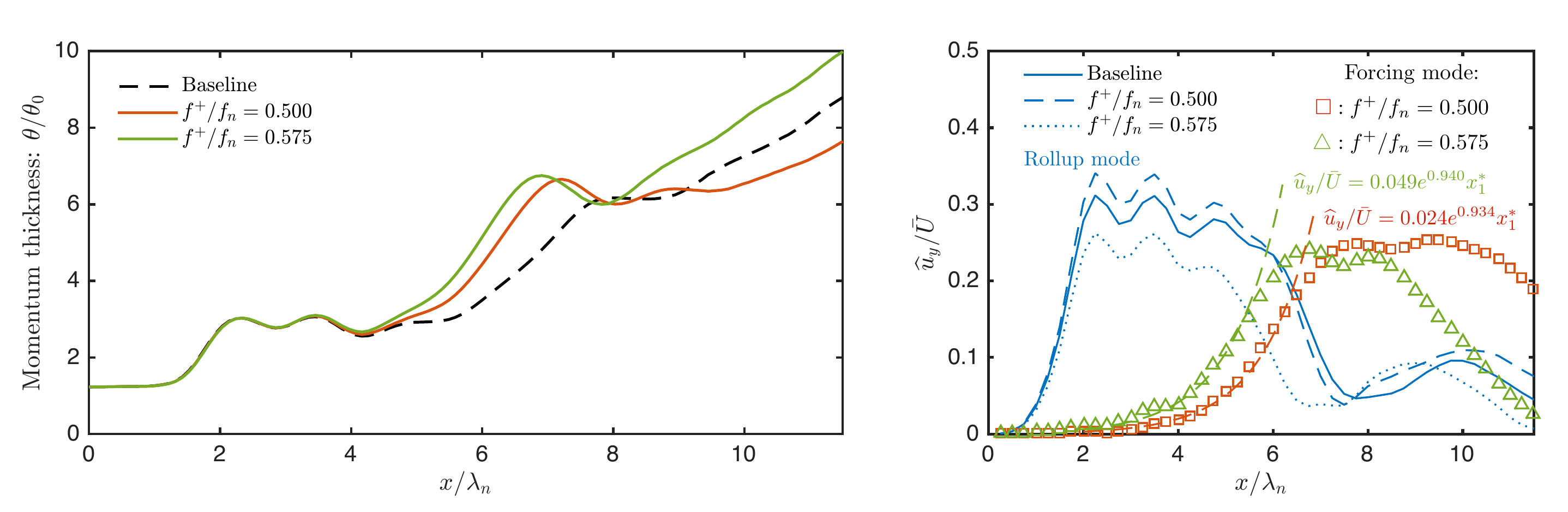}
	\caption{\label{fig:VortexTrackLow2} Momentum thickness growth (left) and modal amplitude evolution (right) for $\theta_0/w = 0.1$ with forcing frequency close to the first subharmonic of $f_n$.}
\end{figure}

We start our detailed discussion on cases with $\theta_0/w = 0.1$.  The baseline and forced flows are compared in figures \ref{fig:VortexTrackLow1} and \ref{fig:VortexTrackLow2}.  It is found that the forcing is able to change the vortex dynamics by encouraging the merging downstream while keeping the momentum thickness unchanged in the region dominated only by fundamental roll-up.  We first track the center of each individual vortex after roll-up using the $Q$-criterion.  By tracking all formed vortices over a time span of $100$ roll-up periods, we visualize the vortex trajectories by black transparent dots and the contour lines for a representative instantaneous spanwise vorticity field in figure \ref{fig:VortexTrackLow1} (left).  Observing the trajectories, it is clear that, as the vortices convect downstream, they spread away from the centerline ($y/\lambda_n = 0$) earlier upstream in the forced flows compared to the baseline case.  This deviation from the centerline serves as an initiation for vortex merging process \citep{Winant74,HoHuang:JFM82}.  While the vortex merging process takes place, the flow passes through a stage where a pair of vortices become vertically aligned, which then results in the steep spatial growth of the momentum thickness.  For example, the controlled case with $f^+/f_n = 0.500$ shows the vertical alignment to take place at $x/\lambda_n \approx 7$.  By observing the momentum thickness in figure \ref{fig:VortexTrackLow2} (left), we find that this steep growth indeed corresponds to where the trajectories demonstrate deviation from the centerline for each case.  With forcing at these two frequencies, the momentum thickness does not exhibit any change to that of the baseline until this second growth takes place farther upstream compared to the baseline.  

A merging criterion is set up such that two vortices are considered to be merged if their instantaneous mutual distance is less than $0.2\lambda_n$.  By setting this threshold , we show the average streamwise location along with the variation (standard deviation) of where merging takes place by the green-shaded regions for each case in figure \ref{fig:VortexTrackLow1} (left).  Again, by comparing the shaded region from the forced cases with the baseline, it is found that the merging process completes upstream with reduced spatial variation.  Observation made from the vortex tracking and mean flow momentum thickness both enable us to draw the conclusion that the forcing is able to change the vortex dynamics by encouraging the merging downstream while keeping the momentum thickness unchanged in the region dominated only by fundamental roll-up.

Since the vortex merging process is dominated by the subharmonics of the roll-up frequency, we now examine the flow over the frequency domain and focus on low frequency components with $f/f_n \le 1$. The transverse velocity spectra along the centerline are shown in figure \ref{fig:VortexTrackLow1} (right) for each corresponding case on the left.  Comparing the forced cases to the baseline case, we find that the forcing mode in each forced case is amplified downstream, which suggests that both control inputs ($f^+/f_n = 0.500$ and $0.575$) are efficiently leveraging the shear layer instability.  Moreover, the growth rates of these two forcing modes are both well predicted by the local stability analysis as previously depicted in figure \ref{fig:BaselineGR}. The energy distribution among the spectra does not show noticeable change from the baseline case until these two forcing modes start to be amplified at $x/\lambda_n \approx 4$, as shown in figure \ref{fig:VortexTrackLow1}.  As discussed in Section \ref{sec:InstabilityAnalysis}, this streamwise station is in the isolated vortex region with linear perturbation growth.  This location also corresponds to where momentum thickness starts to depart from that of the baseline with forcing, as seen in figure \ref{fig:VortexTrackLow2} (left).  The excited low-frequency instability wave induced by the periodic heating accelerates the emergence of the vortex merging process within the shear layer.  This observation agrees with those from the experimental study by \citet{HoHuang:JFM82} and the numerical simulation by \citet{Kourta:AIAAJ1987}.

Next, we shift our attention to the difference between forcing effects with $f^+/f_n = 0.575$ and $0.500$.  In figure \ref{fig:VortexTrackLow1} (left), even though in both forced cases the vortex starts to deviate from centerline earlier than it does in baseline flow, the deviation takes place earlier for the $f^+/f_n = 0.575$ case than the $f^+/f_n = 0.500$ case.  Correspondingly, the merging location is also farther upstream with $f^+/f_n = 0.575$. Moreover, the second growth in momentum thickness is also taking place earlier with $f^+/f_n = 0.575$ as shown in figure \ref{fig:VortexTrackLow2} (left).  Such observations can be explained by the greater growth rate in the forcing mode that both predicted by stability analysis and exhibited in figure \ref{fig:VortexTrackLow2} (right).  The vortex merging process takes place in a more repetitive manner when forcing is introduced at the frequency with higher growth rate, leading to a lower variation in the vortex merging location visualized by the narrower green shaded region for $f^+/f_n = 0.575$. By observing all merging process identified by our criterion and examining all collected flow field snapshots, we find $17\%$ of vortices are not paired with any other vortices with $f^+/f_n = 0.500$, whereas there are only $13\%$ of vortices not going through pairing process $f^+/f_n = 0.575$ in the streamwise window of $x/\lambda_n < 14$.    

After the vortices undergo merging, on the other hand, vortex trajectories exhibit less vertical variation in the subharmonic forced case than they are in the case with $f^+/f_n = 0.575$.  These observations also agrees with those in the development for momentum thickness.   The $\theta/\theta_0$ profiles also appears to be thinner for $x/\lambda_n > 8$ for the case with $f^+/f_n = 0.500$.  By examining the spatial evolution of spectra in figures \ref{fig:VortexTrackLow1} (right) and \ref{fig:VortexTrackLow2} (right), we find that, even though the amplitude decay in roll-up mode takes place upstream compared to the baseline case for both forced cases, the $f^+/f_n = 0.500$ case shows a faster decay rate than $f^+/f_n = 0.575$ case does.  The forcing mode for $f^+/f_n = 0.500$ maintains its high level of modal amplitude while marching downstream, and suppresses other lower frequency modes to grow, as highlighted by the cleaner spectra in the green circled region compared to those of both baseline and $f^+/f_n = 0.575$ cases in figure \ref{fig:VortexTrackLow1} (right). This reduction is due to the lower spillage from the modal energy at the forcing frequency.  Since the growth of other low frequency modes keeps the shear layer to spread downstream, the relatively flat region in momentum thickness in subharmonic forced case can also be attributed to the absence of these lower frequency modes, as shown in figure \ref{fig:VortexTrackLow2} (left).

\begin{figure}
	\begin{center}
		\includegraphics[scale=0.5 ]{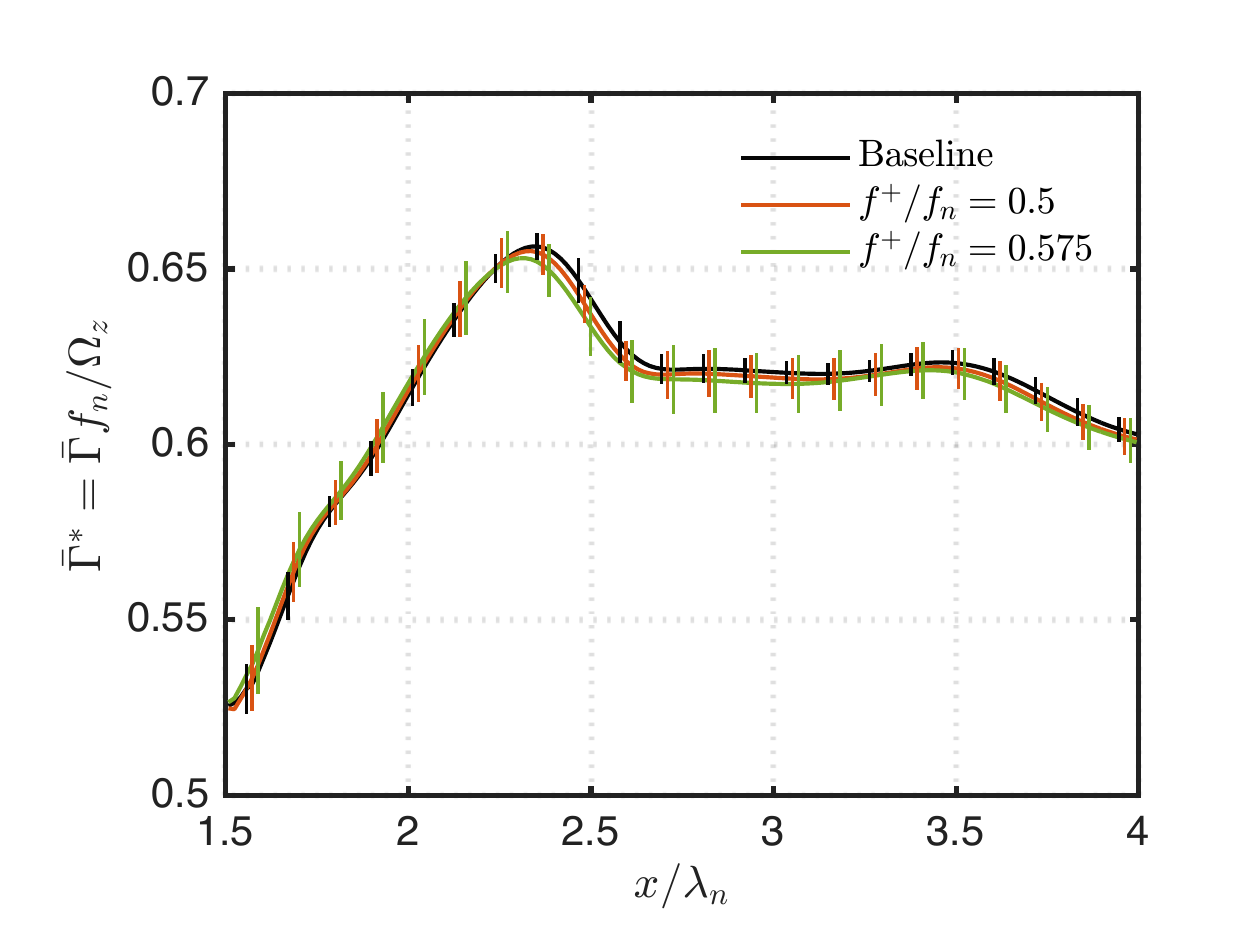}
	\end{center}
	\caption{\label{fig:LowFreqCirc_E2Re4} Normalized ensemble averaged vortex circulation ($\bar{\Gamma}^*$) and its variation (vertical bars) for cases of $\theta_0/w = 0.1$.  We normalized the average circulation, $\bar{\Gamma}$, by the characteristic vorticity that is fed to each vortex in the roll-up period, $\Omega_z/f_n$, from two streams. Low frequency forcing does not significantly change the mean strength but introduces greater variation in the strength of each vortex.}
\end{figure}

As pointed out in section \ref{sec:CtrlMech}, the thermal actuation can produce oscillatory vorticity flux over the trailing edge. This extra fluctuating vorticity generation is in turn fed to each formed vortex and leads to the variation in its strength from one  vortex to another. This is shown in figure \ref{fig:LowFreqCirc_E2Re4}, where normalized circulation at each streamwise station is averaged over each formed vortex.  We observed that, in both forced cases, the mean vortex strength does not significantly change from that of the baseline, but the variation in strength is larger.  The variation in vortex strength can tip over the spatial balance from one vortex to another, resulting in its trajectory deviation from the centerline.  The spread of shear layer can be attributed to both the random trajectory deviation and vortex merging process.  As the strength variation in $f^+/f_n = 0.575$ is greater than $f^+/f_n = 0.500$, the trajectory deviation and merging are indeed taking place earlier for $f^+/f_n = 0.575$ than for $f^+/f_n = 0.500$.

The forcing effects with $f^+/f_n = 0.500$ and $0.575$ in cases of initial momentum thicknesses, $\theta_0/w = 0.05$ and $0.25$ are qualitatively similar to cases of $\theta_0/w = 0.1$ discussed above. Additional discussions for these two $\theta_0/w$ cases are provided in appendix \ref{sec:theta-not025.005}.

\subsection{Excitation of fundamental roll-up ($f^+ \approx f_n$)}
\label{sec:HighFQForcing}

Let us discuss the forcing effects when the forcing frequency is close to $f_n$ with the use of $\dot{q}$ (equation \ref{eq:ZeroMeanForcing}). We consider forcing frequencies of $f^+/f_n = 1.0$, $1.16$ and $1.25$ with $\theta_0/w = 0.1$ and $\theta_0/w = 0.05$.  All cases presented in this section are of the same forcing power of $E^+ = 0.741$.  Although we have also studied forced flow with forcing frequency up to $f^+/f_n = 16$, no significant change in the forced two-dimensional shear layer is observed for cases with $f^+/f_n \ge 2$.  

\begin{figure}
	\begin{center}
		\begin{tabular}{c c}
			\begin{overpic}[scale=0.535]{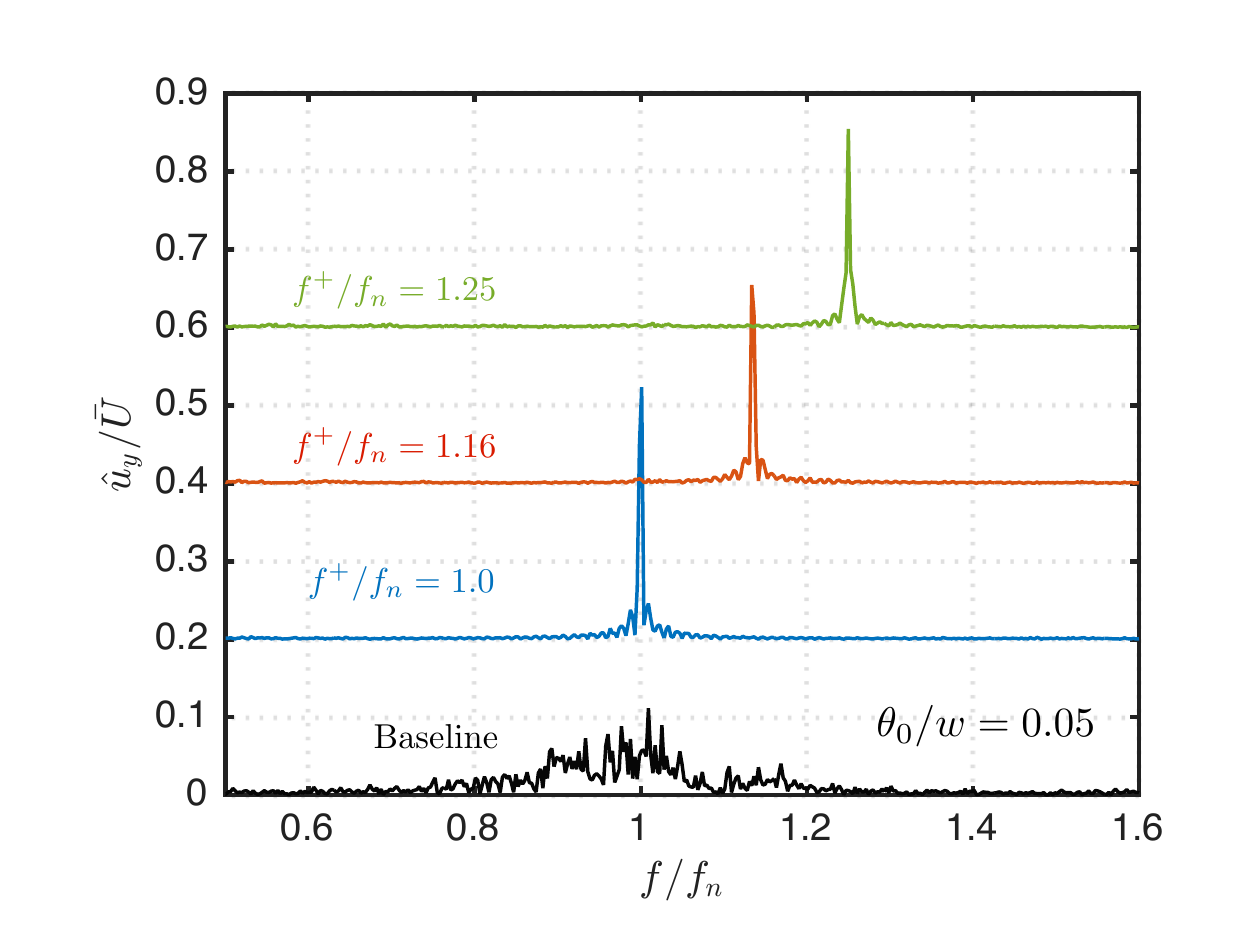}
			\end{overpic} &
			\begin{overpic}[scale=0.535]{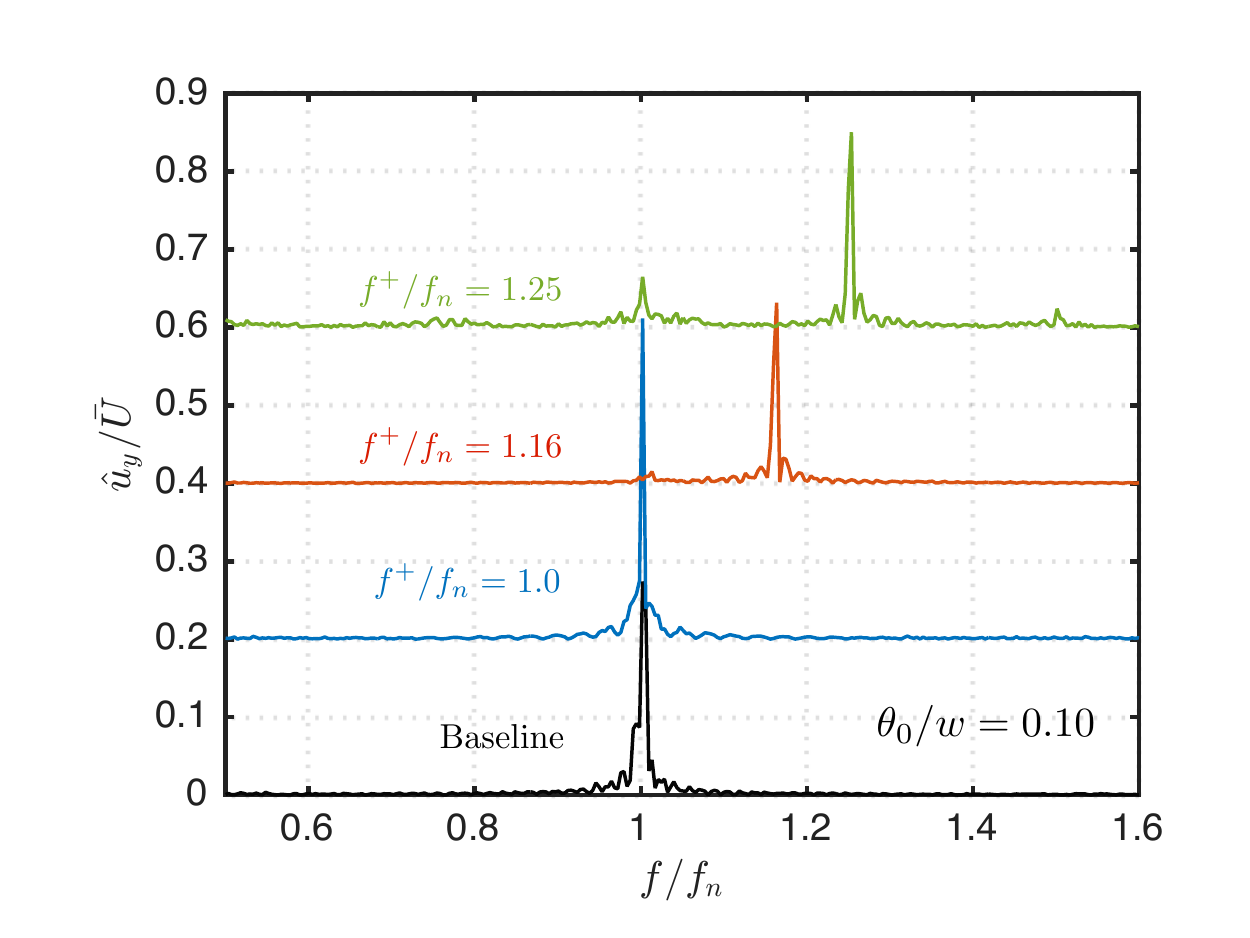}
			\end{overpic}
		\end{tabular}
	\end{center}
	\caption{\label{fig:Lockon} Transverse velocity spectra at $x/\lambda_n = 3.0$: Shedding is observed to lock onto the forcing frequency when $f^+/f_n \in [1.00, 1.25]$. (Left) $\theta_0/w = 0.05$; (Right) $\theta_0/w = 0.1$. The spectra are shifted vertically for graphical clarity.}
\end{figure}

Shown in figure \ref{fig:Lockon} is the transverse velocity spectra for the baseline and forced cases with $f^+/f_n = 1.00$, $1.16$ and $1.25$ at a streamwise station of $x/\lambda_n = 3$ on the centerline. As discussed in Section \ref{sec:BaselineChar}, the stronger wake effect in $\theta_0/w = 0.05$ baseline flow leads to a less synchronized shear layer roll-up and a spectrum with broader distribution of energy across frequencies.  With periodic heating at frequencies considered here, we find that the forcing is able to lock the roll-up onto the forcing frequency, indicated by the prominent peaks at the corresponding forcing frequency for each case for both $\theta_0/w = 0.05$ and $0.1$ in figure \ref{fig:Lockon}. 

\begin{figure}
	\includegraphics[width=1.0\textwidth]{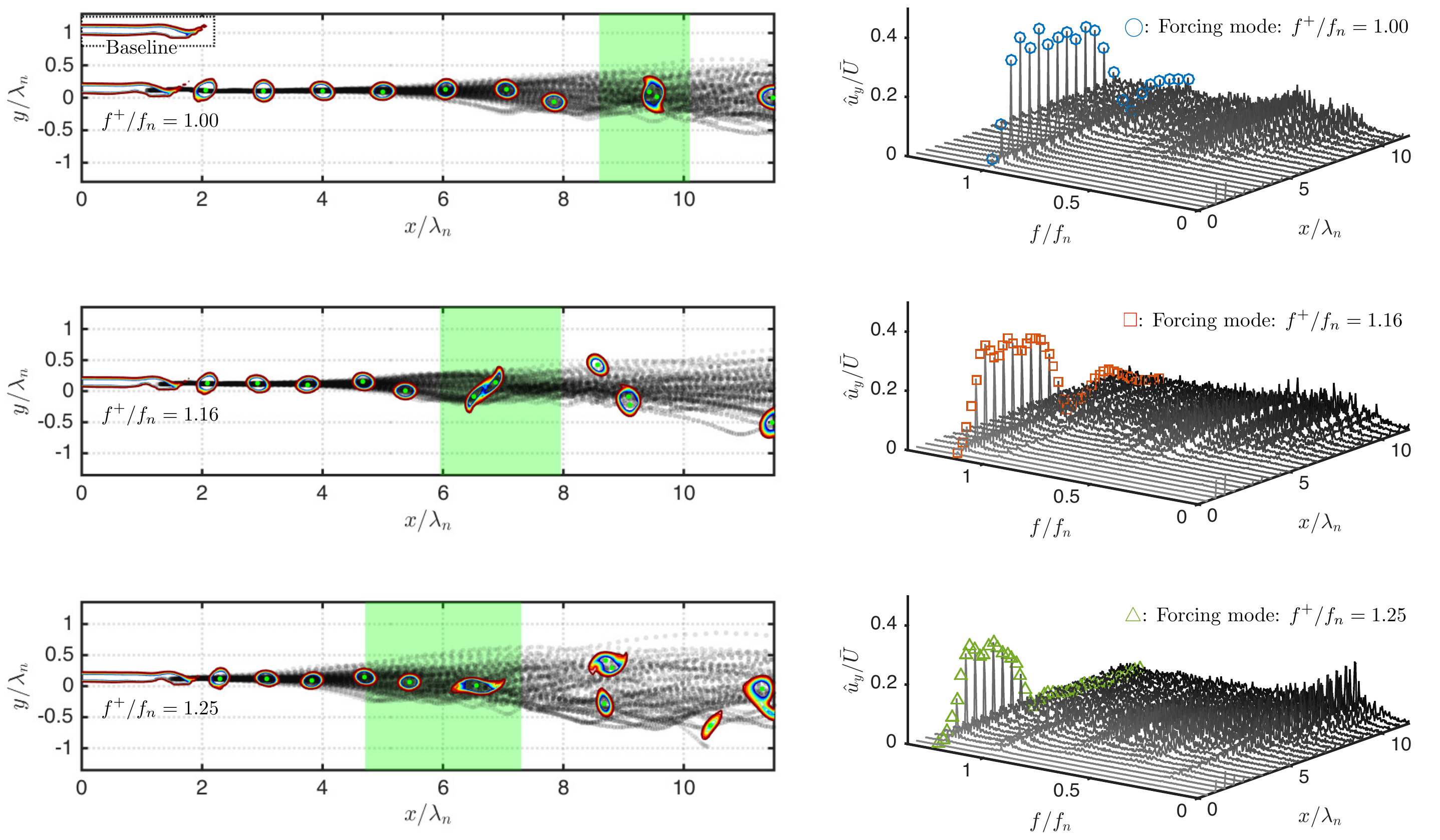}
	\caption{\label{fig:VortexTrackHigh1} Vortex trajectories (left) and modal amplitude growth (right) for $\theta_0/w = 0.1$ cases. With $f^+$ close to $f_n$, the forcing is able to excite the shear layer roll-up and accelerate the roll-up process.}
	\includegraphics[width=1.0\textwidth]{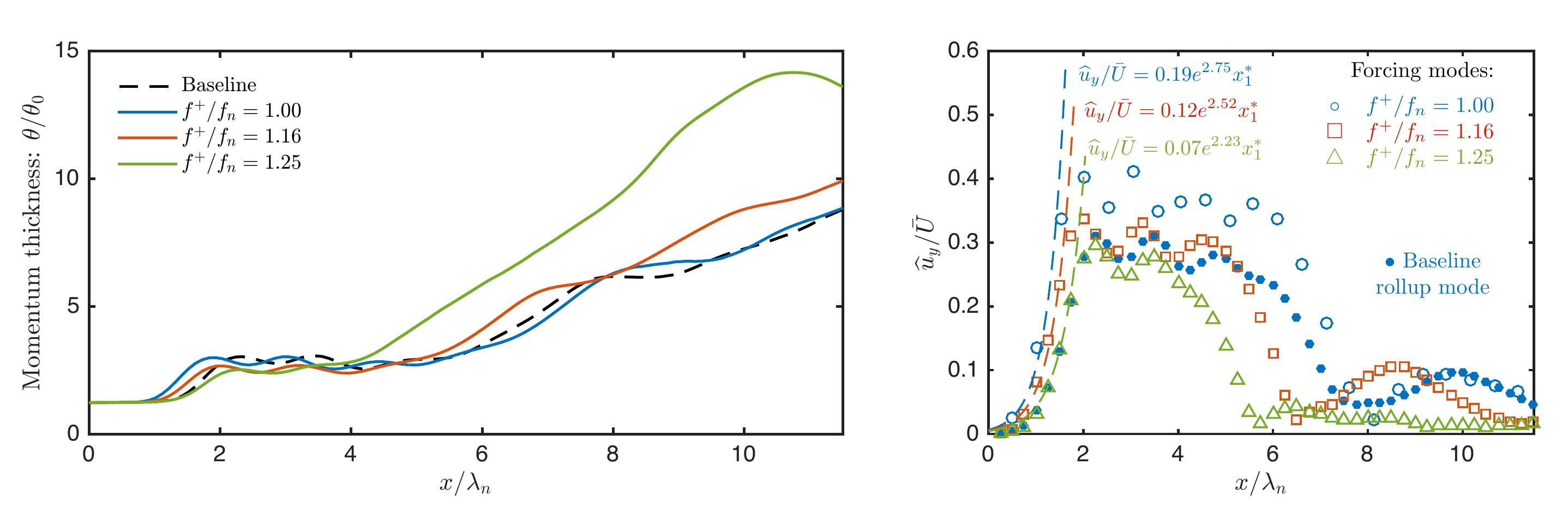}
	\caption{\label{fig:VortexTrackHigh2} Momentum thickness growth (left) and modal amplitude growth (right) for cases with $\theta_0/w = 0.1$ and $f^+/f_n = 1.0$,$1.16$ and $1.25$.}
\end{figure}

Analogous to the discussion for low-frequency forced cases, in figures \ref{fig:VortexTrackHigh1} and \ref{fig:VortexTrackHigh2} we show the vortex trajectories and spatial revolution of transverse velocity spectra and momentum thicknesses for baseline and three forced cases.  The thermal forcing input serves as an excitation to the fundamental instability wave and encourages the shear layer roll-up at the forcing frequency. In cases with $f^+/f_n = 1.0$ and $1.16$, the thermal forcing results in the earlier shedding than that of the baseline. This can be seen by comparing the instantaneous vorticity contours in the initial shedding region: in both cases of $f^+/f_n = 1.0$ and $1.16$, the vorticity sheet formed from $x/\lambda_n = 0$ start to carry streamwise instability wave earlier than the baseline.  Earlier roll-up can also be identified from the first rapid growth of $\theta/\theta_0$ in figure \ref{fig:VortexTrackHigh2} (left).  

\begin{figure}
	\begin{center}
		\includegraphics[scale=0.535]{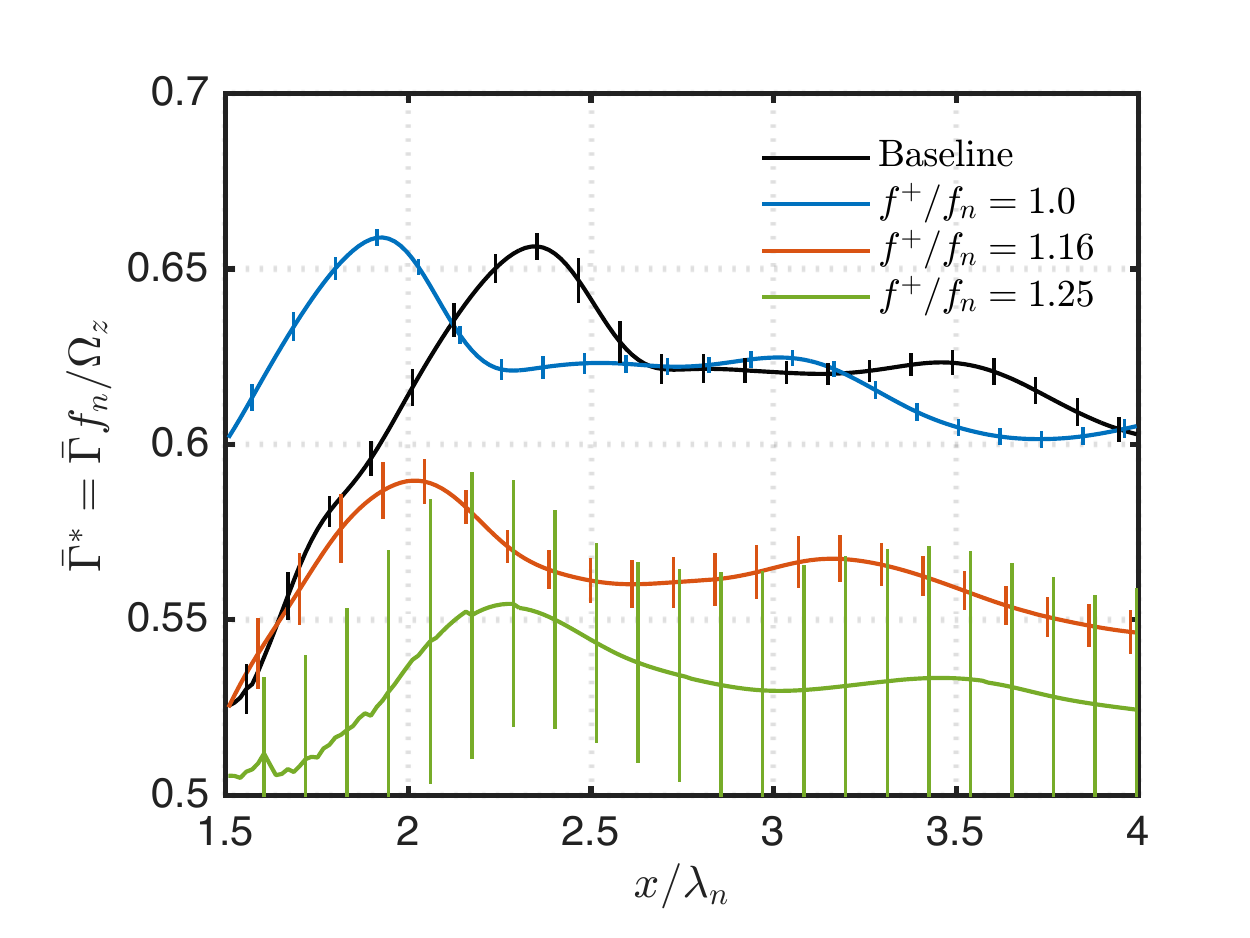}
	\end{center}
	\caption{\label{fig:LowFreqCirc_E2Re4_1} Vortex strength for cases of $\theta_0/w = 0.1$ with $f^+/f_n = 1.0$,$1.16$ and $1.25$. In the control cases where roll-up frequency is locked onto the actuation frequency, the mean strength of the formed vortex is changed accordingly. With greater variation in the strength of each vortex, the enhanced spreading of the shear layer is also observed.}
\end{figure}

With $f^+/f_n = 1.0$, the synchronized shear layer roll-up due to forcing can also be observed from the uniform spacing between each vortex in the vortex train, as show in figure \ref{fig:VortexTrackHigh1} (left).  Also, the forcing is able to suppress the vortex deviating from the centerline.  This leads to an extended isolated vortex region with forcing, suggested by the wider flat region in $1 \le x/\lambda_n \le 5$ in figure \ref{fig:VortexTrackHigh2} (left).  

The spatial development of the frequency spectra and the modal amplitude growth are shown in figures \ref{fig:VortexTrackHigh1} (right) and \ref{fig:VortexTrackHigh2} (right), respectively.  For the three forced cases, the amplitudes of the modes with their frequencies corresponding to the actuation frequencies start to grow immediately behind the trailing edge at $x/\lambda_n = 0$, and their growth rates are well predicted by the local stability analysis, as shown in figure \ref{fig:VortexTrackHigh2} (right).  When the forcing frequency departs farther from $f_n$, the growth rate decreases as observed for $x/\lambda_n \le 2$.  This can be attributed to the instability characteristics of the shear layer. When the forcing frequency is closer to $f_n$, the frequency with the highest spatial growth rate for the perturbation, the forcing effort grows faster since the growth rate is approaching its maximum.  

The frequency spectra of the three forced cases all have distinct peaks at the forcing frequency initially for $x/\lambda_n \le 3$.  As $x/\lambda_n > 3$, lower frequency modes start to grow and the momentum thickness increases accordingly in figure \ref{fig:VortexTrackHigh2} (left). With higher $f^+$, the modal amplitude at the corresponding forcing frequency also decays spatially earlier in $5 \le x/\lambda_n \le 7$.  Earlier decay of the modal amplitude at the forcing frequency also coincides with to the earlier growth of lower frequency modes in the full spectra in figure \ref{fig:VortexTrackHigh1} (right).  Especially for the case of $f^+/f_n = 1.25$, lower frequency modes start to develop farther upstream compared to the baseline flow, resulting in a faster second growth of $\theta/\theta_0$. 

\begin{figure}
	\vspace{0.2in}
	\centerline{\includegraphics[scale=0.535]{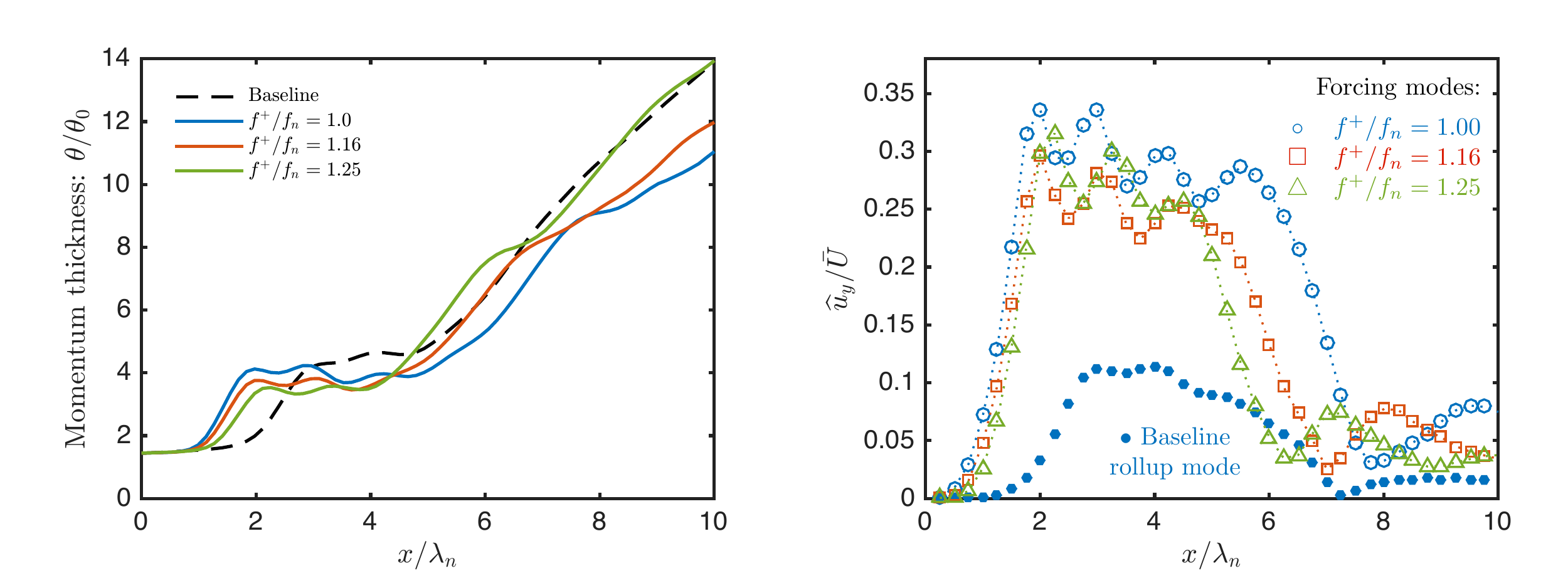}}
	\caption{\label{fig:HighFreqActE2Re2} Momentum thickness growth (left) and modal amplitude growth (right) with high-frequency forcing for the $\theta_0/w = 0.05$ cases. The accelerated roll-up and amplitude growth in the forcing mode are observed to be in qualitative agreement with cases of $\theta_0/w = 0.1$.}
\end{figure}

In the isolated vortex region for the forced cases, thinner $\theta$ is observed when using higher $f^+$.  This can be explained by the weaker vortex strength considering that the total vorticity flux is introduced from the inlet boundary at the same rate. With these compact vortices convecting at the mean shear layer velocity, the higher vortex passage frequency leads to lower amount of vorticity being possessed by the individual vortices, which can be confirmed from figure \ref{fig:LowFreqCirc_E2Re4_1}. For the cases with $f^+/f_n = 1.16$ and $1.25$, the mean vortex strengths are lower than those from the baseline and $f^+/f_n = 1.0$ cases.  With weaker vortices, the reduced mixing of momentum from two streams leads to the thinner momentum thickness in the isolated vortex region.  Also, the strength variations in cases of $f^+/f_n = 1.16$ and $1.25$ are greater than that of the baseline.  This again leads to a faster shear layer spreading than the baseline, similar to the low frequency forced cases.  With $f^+/f_n = 1.0$, on the other hand, the smaller variation in vortex strength leads to a more stable isolated vortex region, leading to the wider streamwise extent than that of the baseline as shown in figure \ref{fig:VortexTrackHigh2} (left).  

For cases with $\theta_0/w = 0.05$ , the lock-on of shear layer roll-up frequency and extended isolated vortex region are still observed when forcing is invoked with $f^+/f_n = 1.00$, $1.16$ and $1.25$, as shown in figure \ref{fig:Lockon} and \ref{fig:HighFreqActE2Re2}.  However, we also note that for $\theta_0/w = 0.25$ cases, no significant changes can be seen in the shear layer for $f^+/f_n = 1.00$, $1.16$, even when using a forcing amplitude of $E^+ = 0.724$, four times of the value considered earlier in this section.

\subsection{Roll-up delay (positive-mean heating)}
\label{sec:PositiveMeanAct}

In this section, forcing effects in the shear layer from using $\dot{q}^p$ (equation \ref{eq:PositiveForcing}) is investigated.  This form of forcing, $\dot{q}^p$, is composed of a positive DC offset added to the oscillatory $\dot{q}$.  The positive DC offset introduces net positive heat transfer from the actuator and raises the mean temperature of the fluid adjacent to the actuator. With the temperature-varying viscosity model discussed in appendix \ref{sec:gov.eqs}, this locally increased temperature leads to a higher viscosity.

\begin{figure}
	\vspace{0.2in}
	\includegraphics[width=1.0\textwidth]{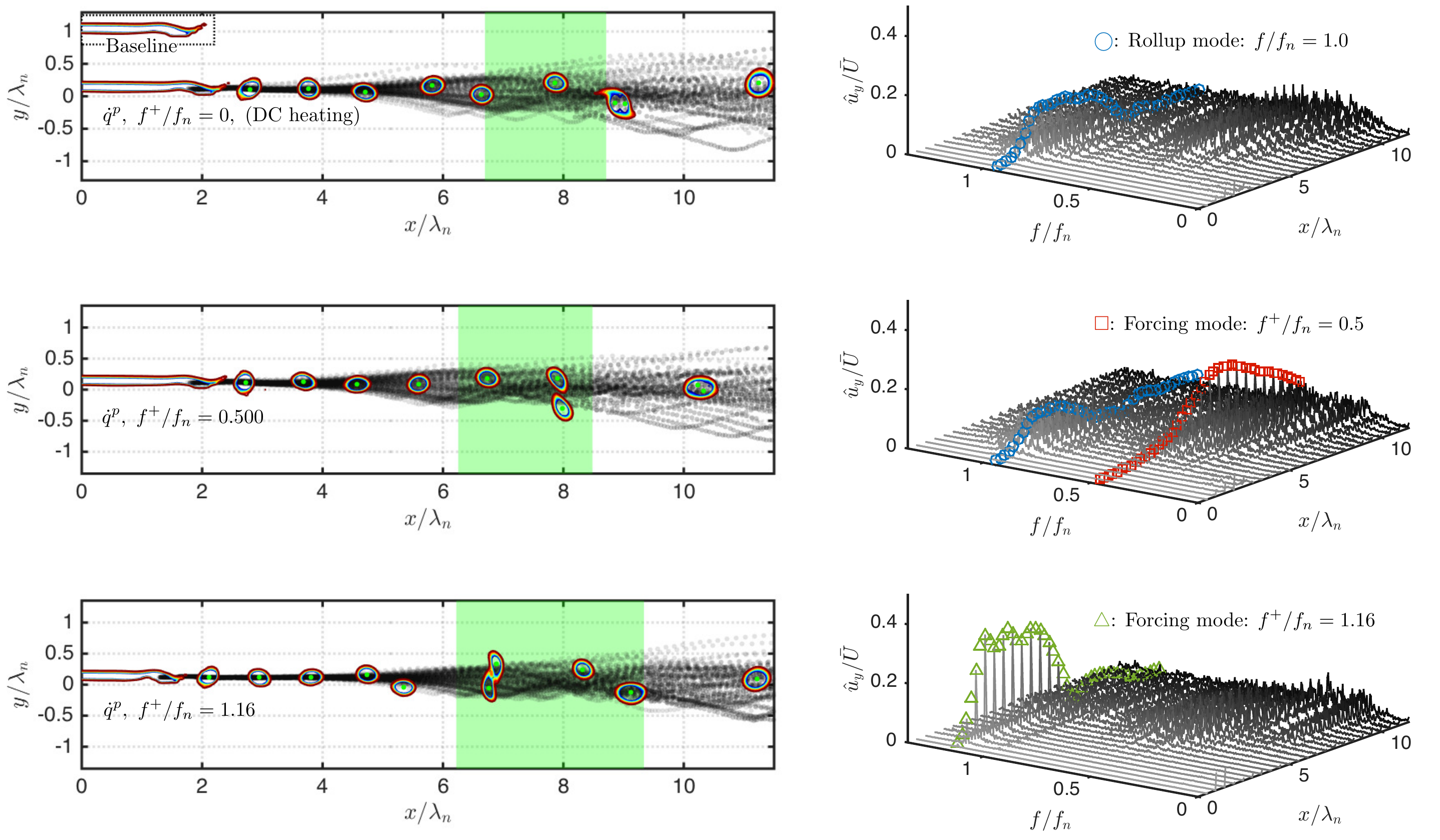}
	\caption{\label{fig:VortexTrack_Posi_1} Vortex trajectories (left) and modal amplitude growth (right) for $\theta_0/w = 0.1$ cases. The merging location and its spatial variation is depicted by the green-shaded region (left) in each case.}
	\vspace{0.2in}
	\includegraphics[width=1.0\textwidth]{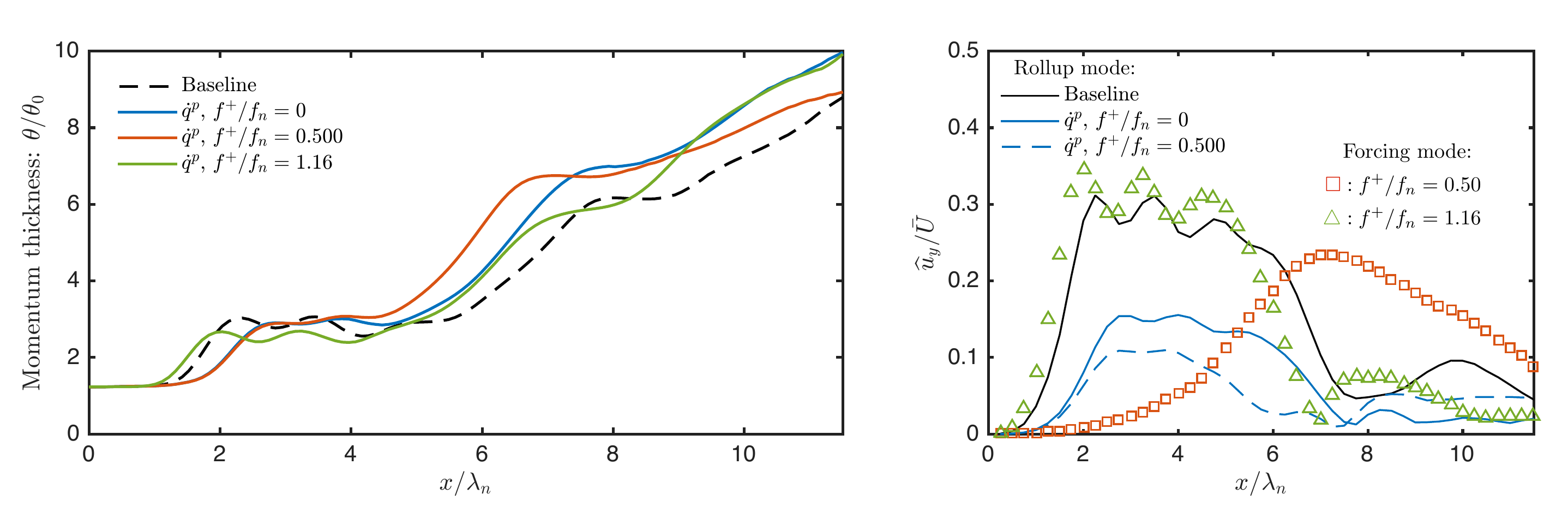}
	\caption{\label{fig:VortexTrack_Posi_2} Momentum thickness growth (left) and modal amplitude growth (right) for low-frequency forcing ($\theta_0/w = 0.1$).}
\end{figure}

\begin{figure}
	\begin{center}
		\includegraphics[scale=0.535]{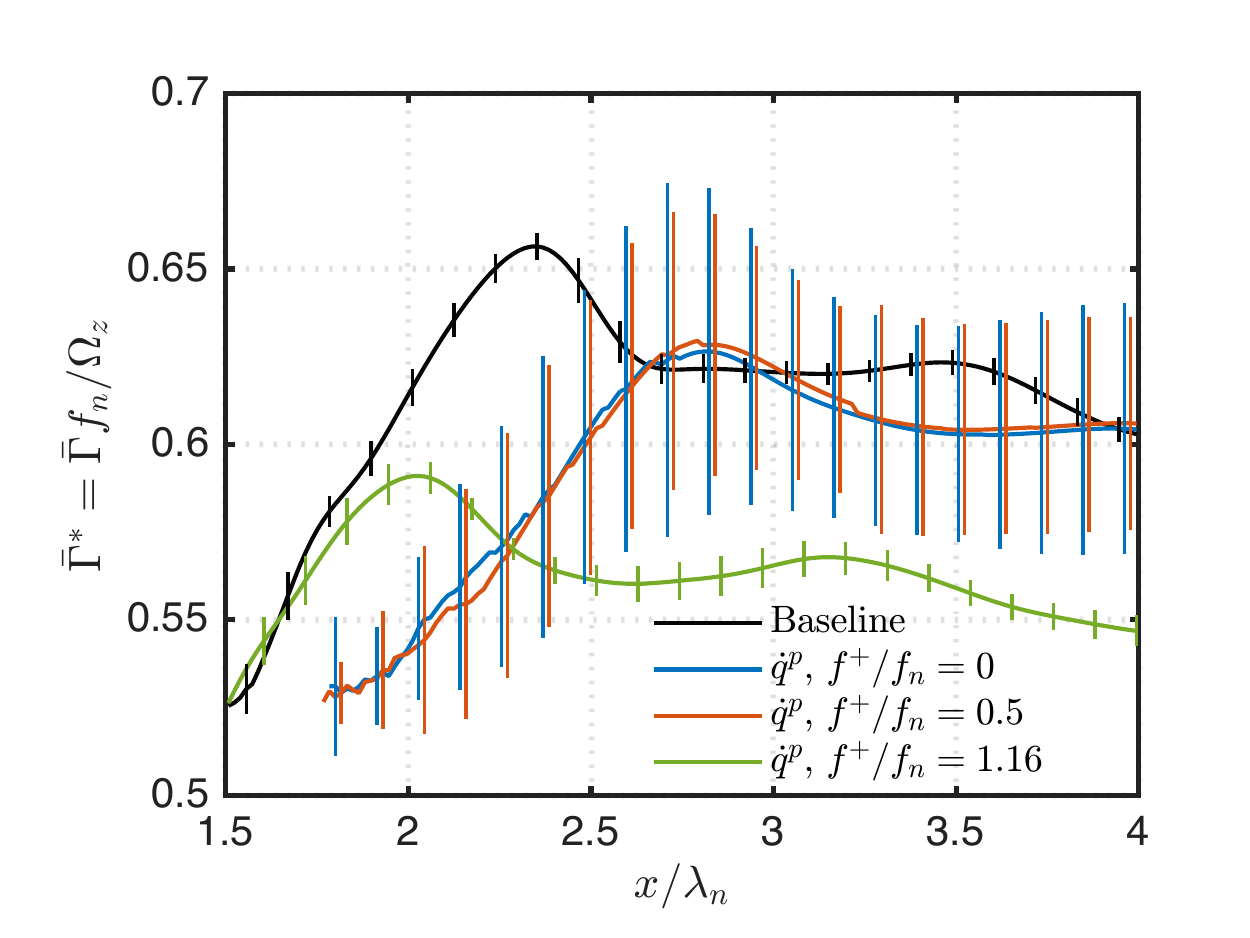}
	\end{center}
	\caption{\label{fig:PosiFreqCirc_E2Re4} Vortex strength for cases of $\theta_0/w = 0.1$.  In the cases with roll-up delay, the vortex strength is observed to have greater variation and encourage the shear layer spreading downstream. All cases are of $\theta_0/w = 0.1$.}
\end{figure}

We consider representative forced cases with $E^+ = 0.741$. The effects of the locally increased viscosity is shown in figure \ref{fig:VortexTrack_Posi_1} (left).  Comparing the cases with $f^+/f_n = 0$ (i.e., constant heating) and $0.500$ to the baseline, the shear layer roll-up is delayed due to the locally increased viscosity.  The delayed shear layer roll-up leads to an elongated recirculation region behind the trailing edge and introduces a stronger wake effect.  As a consequence, the roll-up is observed to be less synchronized, as depicted in figure \ref{fig:VortexTrack_Posi_1} (right) by the less prominent peak in the roll-up mode in the spectra of the DC heating and $f^+/f_n = 0.500$ cases.  The growth in the roll-up mode is also observed to reach a lower maximum amplitude than it does in the baseline flow, as show in figure \ref{fig:VortexTrack_Posi_2} (right). With the stronger wake effect and the less synchronized roll-up, the vortex strength exhibits greater variation, as shown in figure \ref{fig:PosiFreqCirc_E2Re4}.  By the same mechanism discussed in the previous sections, the higher variation leads to the early deviations of the trajectories from the centerline, as well as the early merging of the vortices. Both of these effects cause the shear layer to spread faster downstream in the cases with DC heating and $f^+/f_n = 0.500$.  The delayed roll-up and accelerated spreading can also be observed in the growth of $\theta/\theta_0$ from figure \ref{fig:VortexTrack_Posi_2} (left).  With the subharmonic forcing added, the $f^+/f_n = 0.500$ case exhibits a greater growth in $\theta$ compared to the DC heating case.  In the case with $f^+/f_n = 1.16$, the oscillatory component of $\dot{q}^p$ dominates over the DC component and still accelerates the shear layer roll-up.  The changes in the vortex trajectories (figure \ref{fig:VortexTrack_Posi_1} (left)), vortex strength (figure \ref{fig:PosiFreqCirc_E2Re4}), $\theta/\theta_0$ growth (figure \ref{fig:VortexTrack_Posi_2} (left)) and the forcing mode growth (figure \ref{fig:VortexTrack_Posi_2} (right)) show similar traits to those observed in the case using $\dot{q}$ with $f^+/f_n = 1.16$.

\begin{figure}
	\begin{center}
		\includegraphics[scale=0.535]{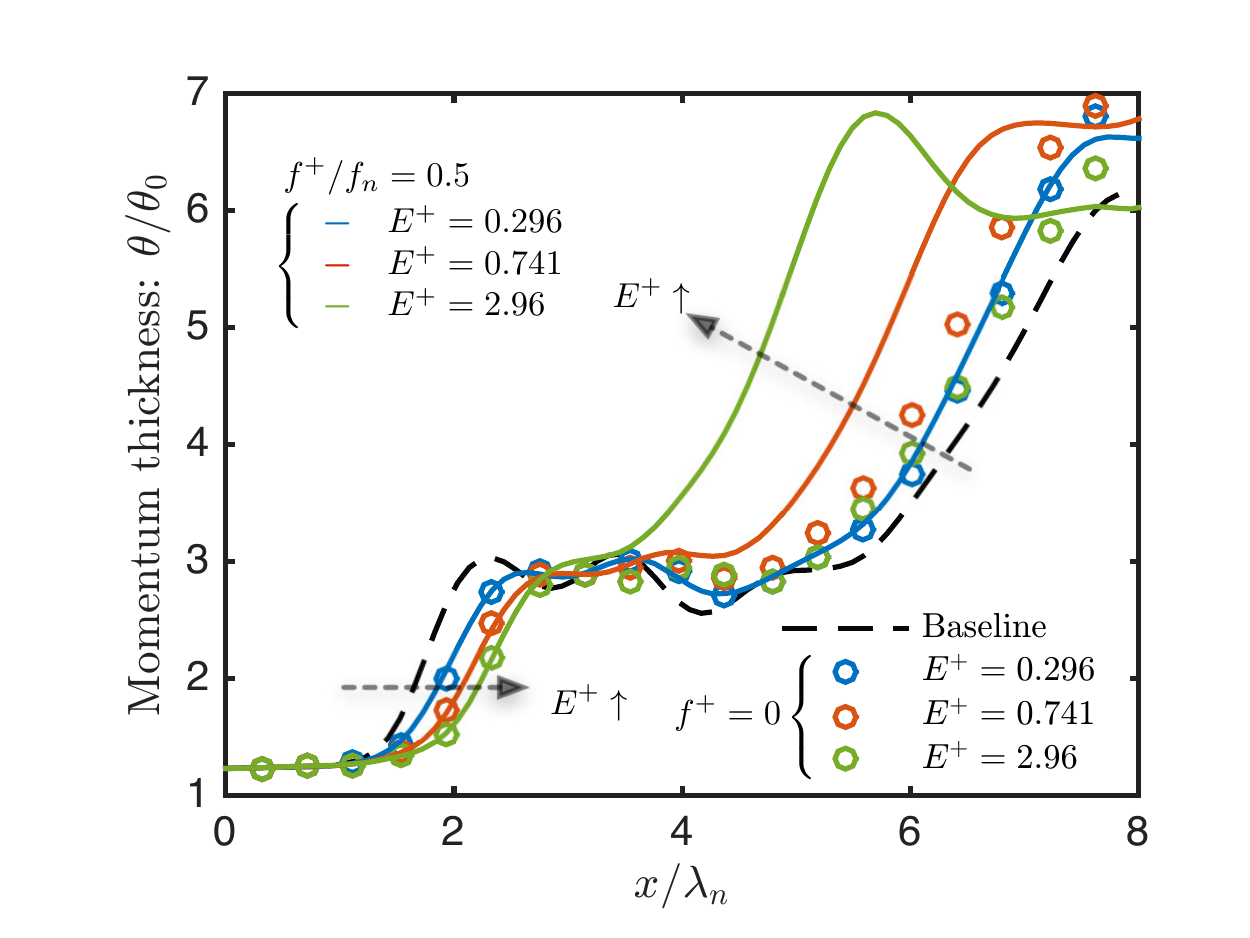}
	\end{center}
	\caption{\label{fig:Posi_ForcingAmp} Effects of forcing amplitude ($E^+$) in the DC heating and subharmonic forced cases with $\dot{q}^p$. All cases are of $\theta_0/w = 0.1$.}
\end{figure}

In the subharmonic forced case with $\dot{q}^p$, the accelerated merging built upon the delayed roll-up draws additional interests.  Motivated by the observation, for cases of DC heating and $f^+/f_n = 0.500$, we further examine the effects of the forcing amplitude, $E^+$, in the growth of $\theta$, as shown in figure \ref{fig:Posi_ForcingAmp}.  We observe that the shear layer roll-up location, suggested by the first growth in $\theta$, is delayed farther downstream with increased $E^+$, in both DC heating and subharmonic forced cases.  Meanwhile, the acceleration in the spreading (second growth of $\theta$) is also enhanced by the subharmonic forcing with increasing $E^+$.  

Comparing the forcing effects of $\dot{q}$ and $\dot{q}^p$, we observe that the oscillatory components from both type impose similar forcing effects to the shear layer.  The positive DC component in $\dot{q}^p$ is able to delay the shear layer roll-up when it is not excited by the oscillatory component.  

\section{Concluding remarks}
\label{sec:conclusion}
\begin{figure}
	\begin{center}      
		\includegraphics[width=0.8\textwidth]{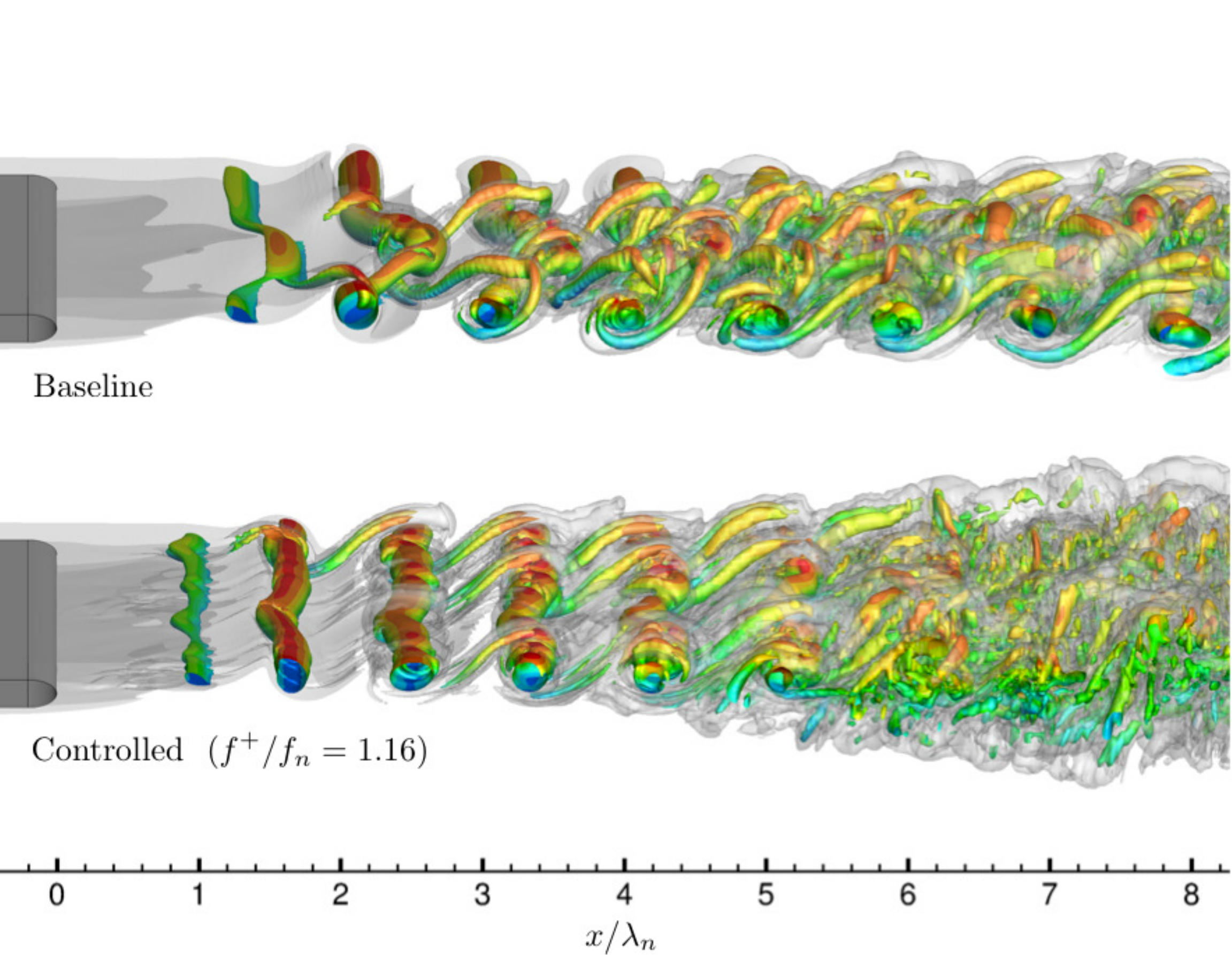}
	\caption{\label{fig:3DMixLayer} 3D LES of shear layers with $\theta_0/w = 0.1$ for the baseline and controlled cases ($f^+/f_n = 1.16$). Flow visualization uses the $Q$-isosurface colored by the streamwise velocity and isosurface of vorticity magnitude overlaid in transparent gray.}
	\end{center}
\end{figure}

We numerically examined the effectiveness of local periodic heating for modifying a spatially developing shear layer downstream of a splitter plate.  The early evolution of the baseline shear layer is characterized by three regions; namely, the region where vortices rolls up, the second region where vortices advect along the centerline, and the third region where vortices deviate from the centerline, leading to the merging process.  The periodic heating from the trailing edge introduces a thermal perturbation to the fluid adjacent to the actuator, and consequently generates oscillatory vorticity flux over its surface and baroclinic torque in its vicinity.  These added sources of vorticity perturb the strength of the vortices formed from the shear layer roll-up, and result in the change in the shear layer dynamics.  

This thermal actuation technique is shown to be capable of serving as an excitation source to both the fundamental and subharmonic instabilities.  When using the forcing frequency close to the first subharmonic of the shear layer roll-up frequency, the forcing can encourage vortices to deviate from the centerline and accelerate the merging process.  When the periodic heating excites the shear layer roll-up, such forcing is able to shift the roll-up frequency to the actuation frequency, and modify the mean strength of each vortex.  The local momentum thickness in the isolated vortex region is also accordingly modified before further spreading takes place.  The thermal actuation with positive-mean is also observed to delay the shear layer roll-up.  In spite of the delayed roll-up, the subharmonic oscillatory component is still able to accelerate the downstream spreading of the shear layer.  We also find that the shear layer spreading rate can be characterized by the synchronized nature of the shear layer roll-up.  The lower level of roll-up synchronization correlates with the greater variance in the vortex strength, less repeatable vortex trajectories, and higher spreading rate of the shear layer. 

As a final note, let us briefly mention that the same control setting is observed to be effective in modifying three-dimensional free shear layers.  We perform large-eddy simulations (LES) using the same setup as in 2D simulations but extend the computational domain in the spanwise direction with the extent of $z/w = [-10, 10]$ with spanwise periodicity.  Shown in figure \ref{fig:3DMixLayer} is the instantaneous flow field, visualized using the isosurface of $Q$-criterion and vorticity magnitude, for the baseline and a controlled case of $\theta_0/w = 0.1$.  For the controlled case, the excitation of shear layer roll-up instability is still effective in modifying the 3D shear layer with $f^+/f_n = 1.16$.  Also, the enhanced shear layer spreading can also be observed in the flow visualization, as indicated by the wide transverse extent of the vortical structure in the controlled case compared to the baseline flow.  While the use of present control approach requires further investigation in 3D turbulent flow, the shown result exhibits promising capabilities of the present actuation technique.  In fact, this type of actuation has been shown to be effective in suppressing turbulent separated flow at a chord-based Reynolds number of $23,000$ over a canonical airfoil in another study by the current authors \citep[see][]{Yeh:AIAA2017}, showing the potential of thermal-based flow actuation in aerodynamic flow control applications.

We demonstrated that the nonlinear dynamics of a spatially developing shear layer can be modified by local oscillatory heat flux as a control input.  This thermal-energy-deposition-based actuation is able to trigger the shear layer instability without relying on mass or momentum injections, and can be considered as a potential candidate for flow control actuation mechanism for a range of applications, especially for those that do not permit internal installation space for the actuators or sizable weight additions.  We believe that this study provides a foundation for flow control using modern thermal-energy-based actuators.

~\\
The authors acknowledge the U.S. Army Research Office for supporting this project (Award Number W911NF-14-1-0224). The computations were supported by the High Performance Computing Modernization Program at the Department of Defense and the Research Computing Center at the Florida State University.

\appendix
\section{Spatial oscillation in momentum thickness over the isolated vortex region}\label{sec:mom.ripple}
\begin{figure}
	\begin{center}
	\includegraphics[scale=0.535]{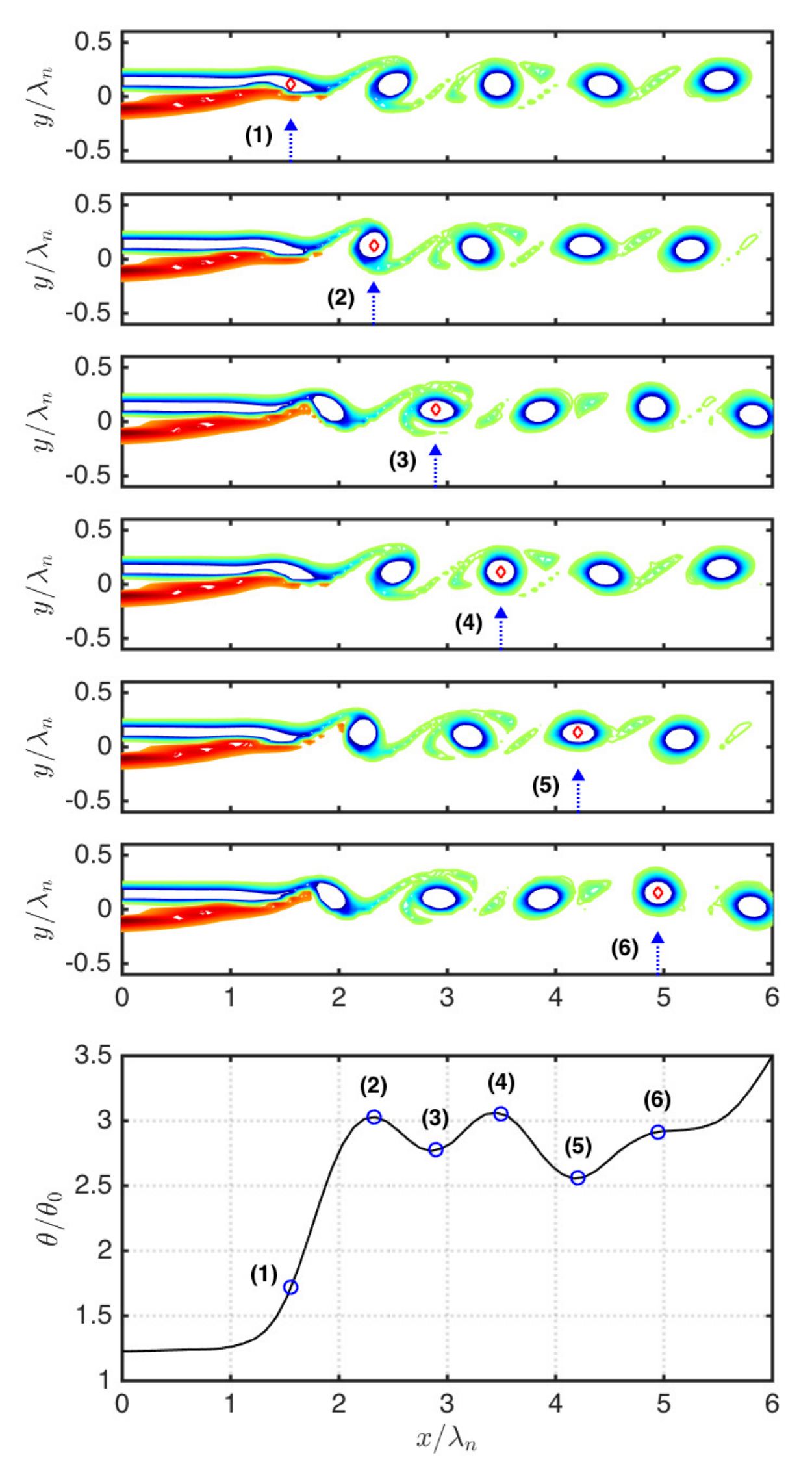}
	\caption{\label{fig:Ripple} The spatial evolution of the vortex orientation.  The spatial oscillation in the $\theta/\theta_0$ profile in the isolated vortex region is caused by the orientation of the elliptic vortices.}
	\end{center}
\end{figure}

This spatial variation in the momentum thickness profile in the isolated vortex region is due to the changing orientation of the elliptic vortices. This spatial evolution in the vortex orientation is illustrated in figure \ref{fig:Ripple} with six snapshots of instantaneous vorticity field (top) and the momentum thickness profile (bottom) with the spatial oscillation.  Note that the flow fields are visualized in equal aspect ratio in both streamwise and transverse directions.  We clearly observe that, at stations (3) and (5), the vortex is horizontally stretched, and the $\theta/\theta_0$ profile exhibits local minimum. On the other hand, at stations (2), (4) and (6), the vortex appears circular in shape and results in thicker $\theta/\theta_0$.

\section{Subharmonic forced cases for $\theta_0/w = 0.05$ and $0.25$}
\label{sec:theta-not025.005}

\begin{figure}
\vspace{1cm}
	\includegraphics[width=1.0\textwidth]{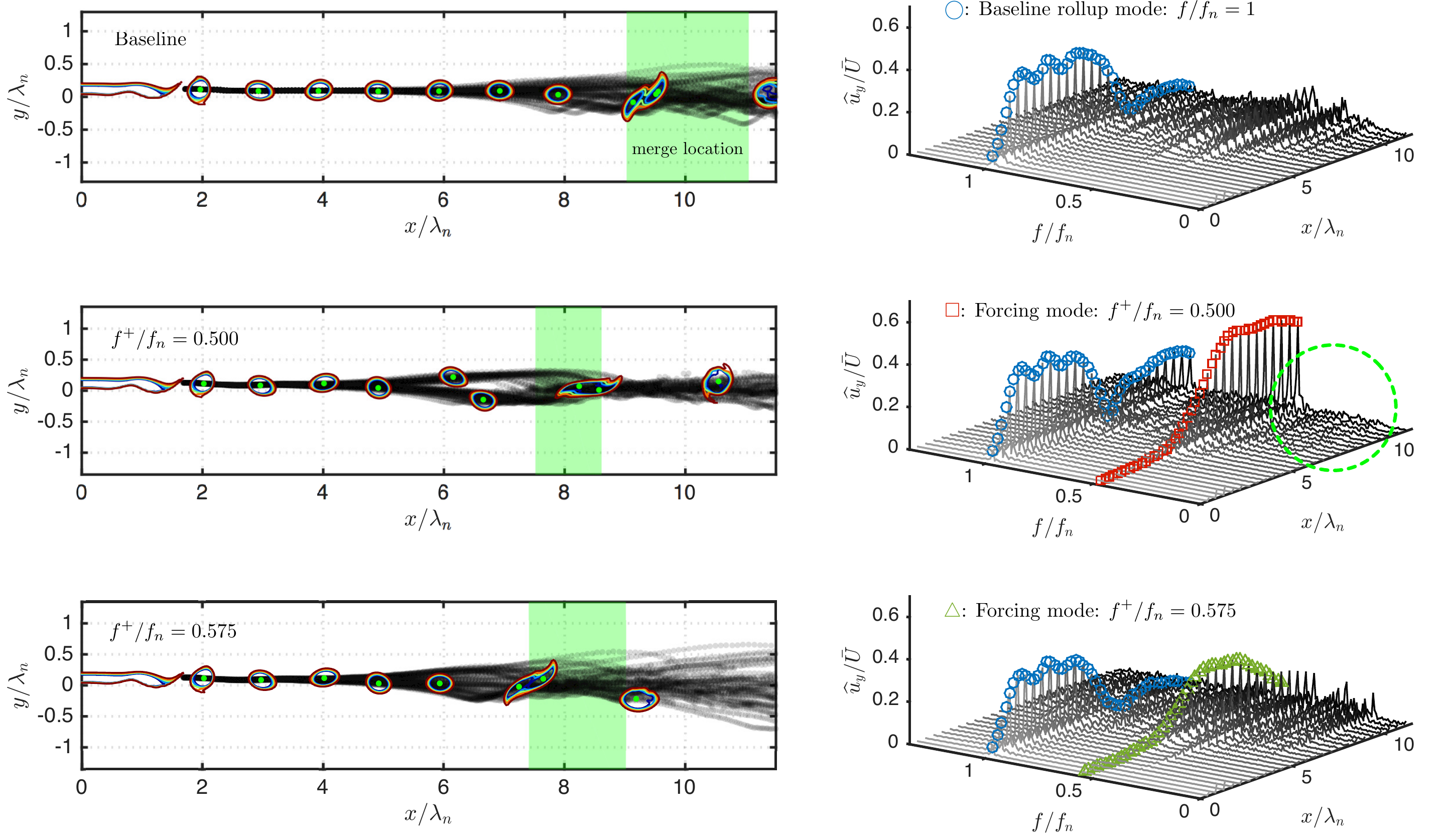}
	\caption{\label{fig:VortexTrackLow_E3Re1} Vortex trajectories (left) and modal amplitude growth (right) for the $\theta_0/w = 0.25$ cases. The merging location and its spatial variation is depicted by the green-shaded region (left) in each case.}
\end{figure}

\begin{figure}
	\begin{center}
		\begin{tabular}{c c}
			\includegraphics[scale=0.535]{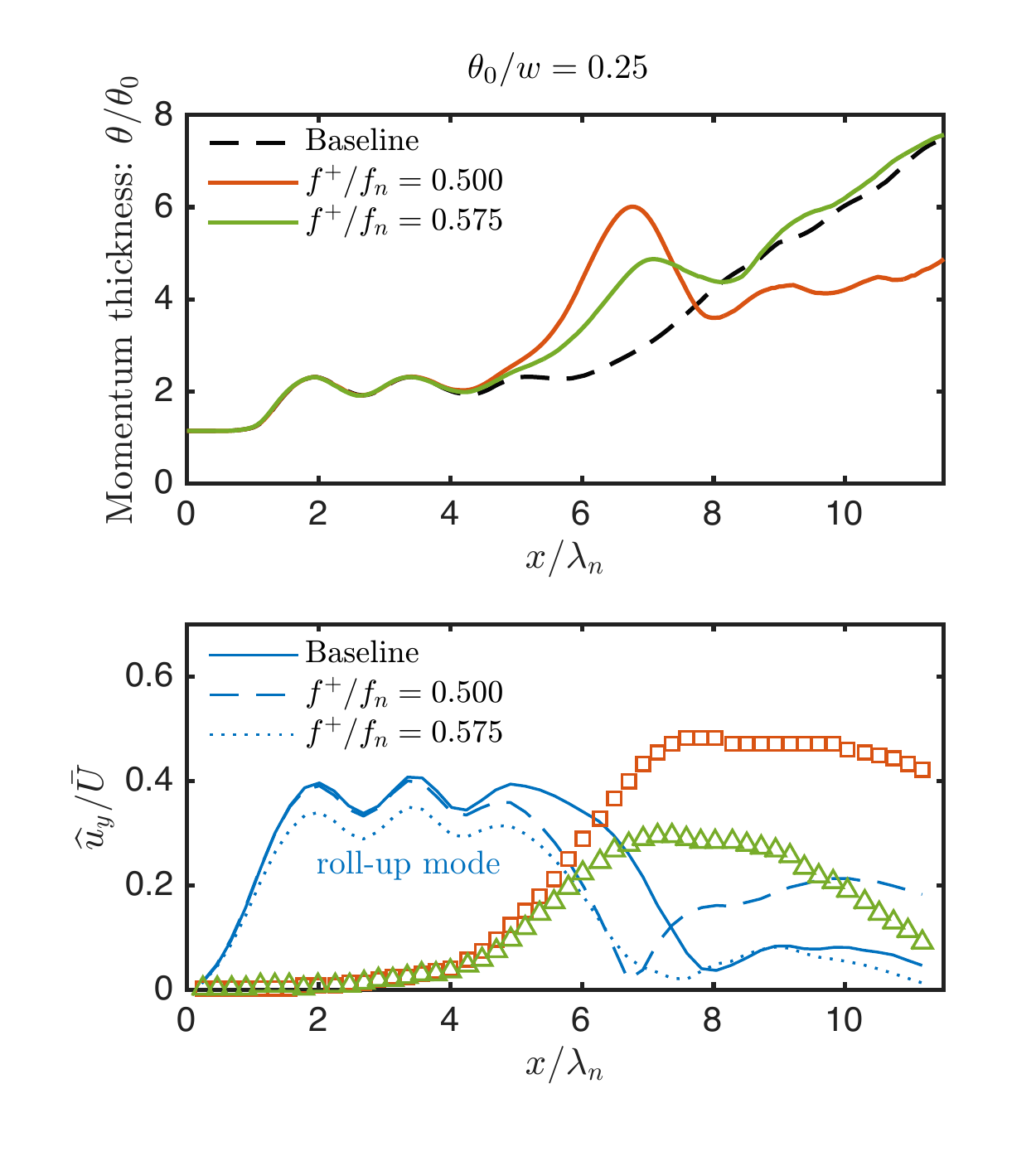} &
			\includegraphics[scale=0.535]{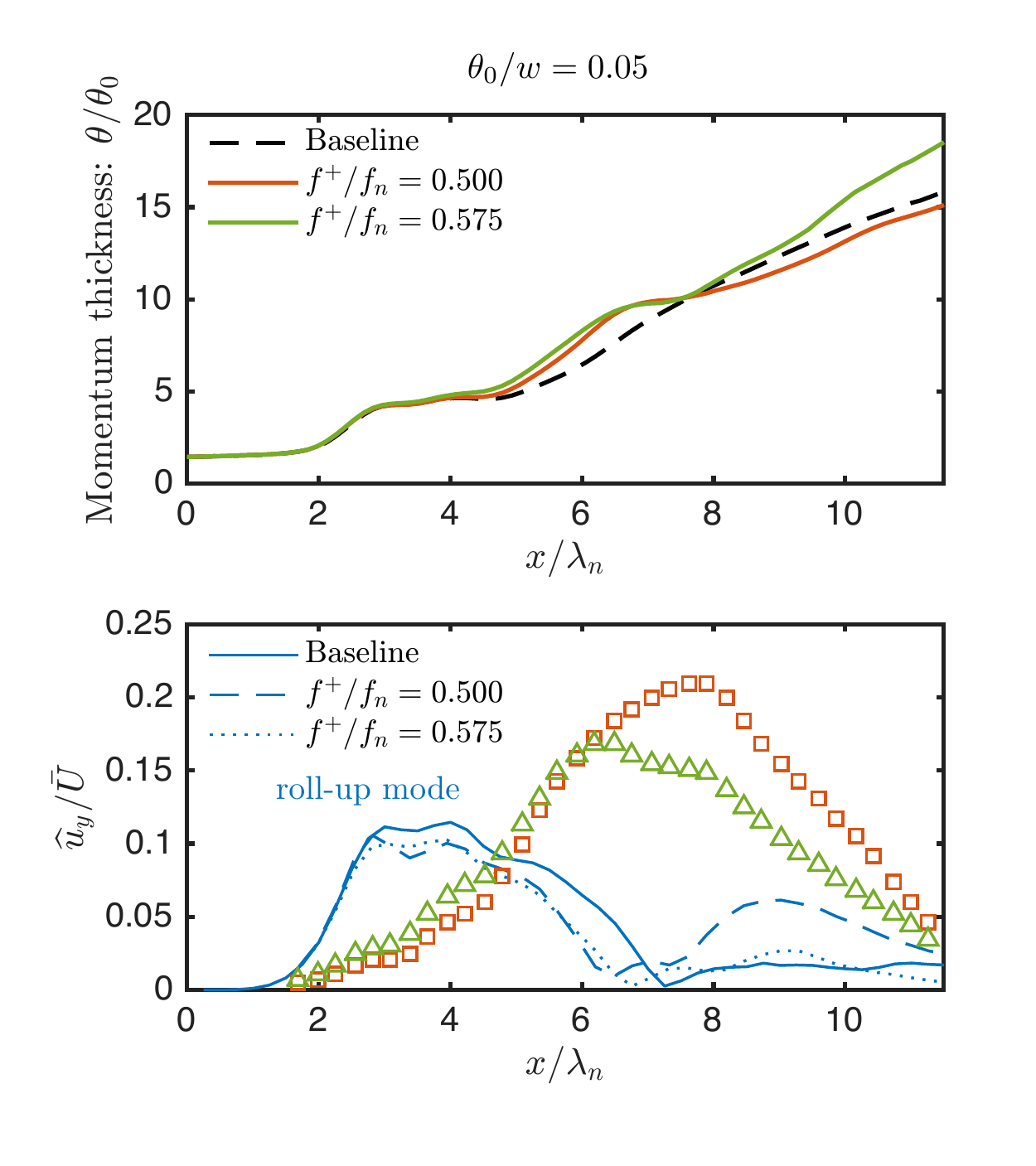}
		\end{tabular}
	\end{center}
	\caption{\label{fig:LowFreqActE2Re2&E3Re1} Momentum thickness growth (top) and modal amplitude growth (bottom) with low-frequency forcing. Cases with $\theta_0/w = 0.25$ (left) and $\theta_0/w = 0.05$ (right). In bottom figures, {\color{red2}$\square$.} represents forcing mode for $f^+/f_n = 0.500$ and {\color{green2}$\circ$} represents $f^+/f_n = 0.575$.}
\end{figure}

We find noteworthy effects with subharmonic forcing using $\hat{q}$ in the case with $\theta_0/w = 0.25$, as shown in figure \ref{fig:VortexTrackLow_E3Re1} (left). Even though the vortex starts to deviate from the centerline earlier in both of the forced cases compared to the baseline, the vortex merging process takes place in a more repetitive (locked) manner when forcing is introduced at the first subharmonic of the roll-up frequency.  The ordered and repetitive trajectories in $f^+/f_n = 0.500$ leaves a white region absent of any trajectories around $x/\lambda_n \approx 6.5$, where no vortices have traveled through.  The use of subharmonic frequency allows for vortex merging by successive pairs along the vortex train.  Also, after merging, vortex trajectories show significant reduction in their variations in the subharmonic forced case than in the $f^+/f_n = 0.575$ case.  The growths of the modal amplitudes is shown in figure \ref{fig:LowFreqActE2Re2&E3Re1} (bottom), where the modal amplitudes at both forcing frequencies are amplified downstream.  The modal amplitude at the forcing frequency in $f^+/f_n = 0.500$ case maintains its high level while marching downstream, and suppresses other lower frequency modes to grow, as highlighted by the clean spectra in the green circled region, compared to the baseline and $f^+/f_n = 0.575$ cases in figure \ref{fig:VortexTrackLow_E3Re1} (right), due to reduced spillage of the modal energy at forcing frequency.  Since the growth of other low frequency modes keeps the shear layer spreading downstream, the slower growth in $\theta$, as shown in figure \ref{fig:LowFreqActE2Re2&E3Re1} (top) for both $\theta_0$ cases, can also be attributed to the absence of these lower frequency modes.  

\bibliographystyle{jfm}
\bibliography{CAYeh_All}

\begin{thebibliography}{51}
\expandafter\ifx\csname natexlab\endcsname\relax\def\natexlab#1{#1}\fi
\def\au#1{#1} \def\ed#1{#1} \def\yr#1{#1}\def\at#1{#1}\def\jt#1{\textit{#1}}
  \def\bt#1{#1}\def\bvol#1{\textbf{#1}} \def\vol#1{#1} \def\pg#1{#1}
  \def\publ#1{#1}\def\arxiv#1{#1}\def\org#1{#1}\def\st#1{\textit{#1}}

\bibitem[Abe {\em et~al.\/}(2008)Abe, Takizawa, Sato \& Kimura]{Abe:AIAAJ2008}
{\sc \au{Abe, T.}, \au{Takizawa, Y.}, \au{Sato, S.} \& \au{Kimura, N.}}
  \yr{2008}  \at{Experimental study for momentum transfer in a dielectric
  barrier discharge plasma actuator}.  \jt{AIAA Journal}  \bvol{46}~(9),
  \pg{2248--2256}.

\bibitem[Adamovich {\em et~al.\/}(2012)Adamovich, Little, Nishihara, Takashima
  \& Samimy]{Adamovich:AIAA2012}
{\sc \au{Adamovich, I.~V.}, \au{Little, J.}, \au{Nishihara, M.}, \au{Takashima,
  K.} \& \au{Samimy, M.}} \yr{2012} Nanosecond pulse surface discharges for
  high-speed flow control.  \bt{In {\em AIAA Paper 2012-3137\/}}.

\bibitem[Akins {\em et~al.\/}(2015)Akins, Singh \&
  Little]{AkinsLittle:AIAA2015}
{\sc \au{Akins, D.}, \au{Singh, A.} \& \au{Little, J.}} \yr{2015} Effects of
  pulse energy on shear layer control using surface plasma discharges.  \bt{In
  {\em AIAA Paper 2015-3344\/}}.

\bibitem[Aleksandrov {\em et~al.\/}(2010)Aleksandrov, Kindysheva, Nudnova \&
  Starikovskiy]{Aleksandrov:JPD2010}
{\sc \au{Aleksandrov, N.~L.}, \au{Kindysheva, S.~V.}, \au{Nudnova, M.~M.} \&
  \au{Starikovskiy, A.~Yu}} \yr{2010}  \at{Mechanism of ultra-fast heating in a
  non-equilibrium weakly ionized air discharge plasma in high electric fields}.
   \jt{Journal of Physics D: Applied Physics}  \bvol{43}~(25),  \pg{255201}.

\bibitem[Arnold \& Crandall(1917)]{Arnold:PRB17}
{\sc \au{Arnold, H.~D.} \& \au{Crandall, I.~B.}} \yr{1917}  \at{The thermophone
  as a precision source of sound}.  \jt{Physical Review}  \bvol{10}~(1),
  \pg{22}.

\bibitem[Barone \& Lele(2005)]{BaroneLele:JFM2005}
{\sc \au{Barone, M.~F.} \& \au{Lele, S.~K.}} \yr{2005}  \at{Receptivity of the
  compressible mixing layer}.  \jt{Journal of Fluid Mechanics}  \bvol{540},
  \pg{301--335}.

\bibitem[Bechert(1988)]{Bechert:JFM1988_A}
{\sc \au{Bechert, D.~W.}} \yr{1988}  \at{Excitation of instability waves in
  free shear layers. part 1. theory}.  \jt{Journal of Fluid Mechanics}
  \bvol{186},  \pg{47--62}.

\bibitem[Bechert \& Stahl(1988)]{Bechert:JFM1988_B}
{\sc \au{Bechert, D.~W.} \& \au{Stahl, B}} \yr{1988}  \at{Excitation of
  instability waves in free shear layers. part 2. experiments}.  \jt{Journal of
  Fluid Mechanics}  \bvol{186},  \pg{63--84}.

\bibitem[Bin {\em et~al.\/}(2015)Bin, Oates \& Taira]{Bin:JAP2015}
{\sc \au{Bin, Jonghoon}, \au{Oates, William~S} \& \au{Taira, Kunihiko}}
  \yr{2015}  \at{Thermoacoustic modeling and uncertainty analysis of
  two-dimensional conductive membranes}.  \jt{Journal of Applied Physics}
  \bvol{117}~(6),  \pg{064506}.

\bibitem[Br{\`e}s {\em et~al.\/}(2017)Br{\`e}s, Ham, Nichols \&
  Lele]{Bres:AIAAJ2017}
{\sc \au{Br{\`e}s, G.~A.}, \au{Ham, F.~E.}, \au{Nichols, J.~W.} \& \au{Lele,
  S.~K.}} \yr{2017}  \at{Unstructured large-eddy simulations of supersonic
  jets}.  \jt{AIAA Journal} .

\bibitem[Brown \& Roshko(1974)]{BrownRoshko:JFM1974}
{\sc \au{Brown, G.~L} \& \au{Roshko, A.}} \yr{1974}  \at{On density effects and
  large structure in turbulent mixing layers}.  \jt{Journal of Fluid Mechanics}
   \bvol{64}~(04),  \pg{775--816}.

\bibitem[Cattafesta {\em et~al.\/}(2008)Cattafesta, Song, Williams, Rowley \&
  Alvi]{Cattafesta:PAS08}
{\sc \au{Cattafesta, L.~N.}, \au{Song, Q.}, \au{Williams, D.~R.}, \au{Rowley,
  C.~W.} \& \au{Alvi, F.~S.}} \yr{2008}  \at{Active control of flow-induced
  cavity oscillations}.  \jt{Progress in Aerospace Sciences}  \bvol{44},
  \pg{479--502}.

\bibitem[Cheung \& Lele(2009)]{Cheung:PoF09}
{\sc \au{Cheung, L.~C.} \& \au{Lele, S.~K.}} \yr{2009}  \at{The dynamics of
  nonlinear instability waves in laminar heated and unheated compressible
  mixing layers}.  \jt{Physics of Fluids}  \bvol{21}~(9),  \pg{094103}.

\bibitem[Choi \& Moin(2012)]{ChoiMoin:POF2012}
{\sc \au{Choi, H.} \& \au{Moin, P.}} \yr{2012}  \at{Grid-point requirements for
  large eddy simulation: Chapman?s estimates revisited}.  \jt{Physics of
  Fluids}  \bvol{24}~(1),  \pg{011702}.

\bibitem[Clemens \& Mungal(1995)]{ClemensMungal:JFM1995}
{\sc \au{Clemens, N.~T.} \& \au{Mungal, M.~G.}} \yr{1995}  \at{Large-scale
  structure and entrainment in the supersonic mixing layer}.  \jt{Journal of
  Fluid Mechanics}  \bvol{284},  \pg{171--216}.

\bibitem[Corke {\em et~al.\/}(2010)Corke, Enloe \& Wilkinson]{Corke:ARFM10}
{\sc \au{Corke, T.~C.}, \au{Enloe, C.~L.} \& \au{Wilkinson, S.~P.}} \yr{2010}
  \at{Dielectric barrier discharge plasma actuators for flow control}.
  \jt{Annual Review of Fluid Mechanics}  \bvol{42},  \pg{505--529}.

\bibitem[Corke {\em et~al.\/}(2009)Corke, He \& Patel]{Corke:JA09}
{\sc \au{Corke, T.~C.}, \au{He, C.} \& \au{Patel, M.}} \yr{2009}  \at{Plasma
  flaps and slats: an application of weakly-ionized plasma actuators}.
  \jt{Journal of Aircraft}  \bvol{46}~(3),  \pg{864--873}.

\bibitem[Crighton(1985)]{Crighton:ARFM1985}
{\sc \au{Crighton, D.~G.}} \yr{1985}  \at{The {K}utta condition in unsteady
  flow}.  \jt{Annual Review of Fluid Mechanics}  \bvol{17}~(1),  \pg{411--445}.

\bibitem[Elliott \& Samimy(1990)]{ElliottSamimy:PoF1990}
{\sc \au{Elliott, G.~S.} \& \au{Samimy, M.}} \yr{1990}  \at{Compressibility
  effects in free shear layers}.  \jt{Physics of Fluids}  \bvol{2}~(7),
  \pg{1231--1240}.

\bibitem[Freund(1997)]{Freund:AIAAJ97}
{\sc \au{Freund, J.~B.}} \yr{1997}  \at{Proposed inflow/outflow boundary
  condition for direct computation of aerodynamic sound}.  \jt{AIAA Journal}
  \bvol{35}~(4),  \pg{740--742}.

\bibitem[Garnier {\em et~al.\/}(2009)Garnier, Adams \& Sagaut]{Garnier:2009LES}
{\sc \au{Garnier, E.}, \au{Adams, N.} \& \au{Sagaut, P.}} \yr{2009} {\em Large
  eddy simulation for compressible flows\/}.  \publ{Springer Science \&
  Business Media}.

\bibitem[Glezer \& Amitay(2002)]{Glezer:ARFM2002}
{\sc \au{Glezer, A.} \& \au{Amitay, M.}} \yr{2002}  \at{Synthetic jets}.
  \jt{Annual Review of Fluid Mechanics}  \bvol{34}~(1),  \pg{503--529}.

\bibitem[Greenblatt {\em et~al.\/}(2006)Greenblatt, Paschal, Yao \&
  Harris]{Greenblatt:AIAAJ2006_P2}
{\sc \au{Greenblatt, D.}, \au{Paschal, K.~B.}, \au{Yao, C.-S.} \& \au{Harris,
  J.}} \yr{2006}  \at{Experimental investigation of separation control part 2:
  Zero mass-flux oscillatory blowing}.  \jt{AIAA Journal}  \bvol{44}~(12),
  \pg{2831--2845}.

\bibitem[Gad-el Hak \& Bushnell(1991)]{Gad-el-Hak:JFE1991}
{\sc \au{Gad-el Hak, M.} \& \au{Bushnell, D.~M.}} \yr{1991}  \at{Separation
  control: review}.  \jt{Journal of Fluids Engineering}  \bvol{113}~(1),
  \pg{5--30}.

\bibitem[Ho \& Huang(1982)]{HoHuang:JFM82}
{\sc \au{Ho, C.-M.} \& \au{Huang, L.-S.}} \yr{1982}  \at{Subharmonics and
  vortex merging in mixing layers}.  \jt{Journal of Fluid Mechanics}
  \bvol{119},  \pg{443--473}.

\bibitem[Ho \& Huerre(1984)]{HoHuerre:AR84}
{\sc \au{Ho, C.-M.} \& \au{Huerre, P.}} \yr{1984}  \at{Perturbed free shear
  layers}.  \jt{Annual Review of Fluid Mechanics}  \bvol{16}~(1),
  \pg{365--422}.

\bibitem[Hornung(1989)]{Hornung:Melbourn1989}
{\sc \au{Hornung, H.}} \yr{1989} Vorticity generation and transport.  \bt{In
  {\em 10th Australasian fluid mechanics conference, Paper KS-3\/}}.

\bibitem[Khalighi {\em et~al.\/}(2011{\natexlab{{\em a\/}}})Khalighi, Ham,
  Moin, Lele, Schlinker, Reba \& J.]{Khalighi:ASME2011}
{\sc \au{Khalighi, Y.}, \au{Ham, F.}, \au{Moin, P.}, \au{Lele, S.},
  \au{Schlinker, R.}, \au{Reba, R.} \& \au{J., Simonich}}
  \yr{2011{\natexlab{{\em a\/}}}}  \bt{Noise prediction of pressure-mismatched
  jets using unstructured large eddy simulation}. {Proceedings of ASME Turbo
  Expo}, Vancouver.

\bibitem[Khalighi {\em et~al.\/}(2011{\natexlab{{\em b\/}}})Khalighi, Nichols,
  Ham, Lele \& Moin]{Khalighi:AIAA11}
{\sc \au{Khalighi, Y.}, \au{Nichols, J.~W.}, \au{Ham, F.}, \au{Lele, S.~K.} \&
  \au{Moin, P.}} \yr{2011{\natexlab{{\em b\/}}}} Unstructured large eddy
  simulation for prediction of noise issued from turbulent jets in various
  configurations.  \bt{In {\em AIAA Paper 2011-2886\/}}.

\bibitem[Kourta {\em et~al.\/}(1987)Kourta, Braza, Chassaing \&
  Haminh]{Kourta:AIAAJ1987}
{\sc \au{Kourta, A.}, \au{Braza, M.}, \au{Chassaing, P.} \& \au{Haminh, H.}}
  \yr{1987}  \at{Numerical analysis of a natural and excited two-dimensional
  mixing layer}.  \jt{AIAA Journal}  \bvol{25}~(2),  \pg{279--286}.

\bibitem[Laizet {\em et~al.\/}(2010)Laizet, Lardeau \&
  Lamballais]{Laizet:PoF2010}
{\sc \au{Laizet, S.}, \au{Lardeau, S.} \& \au{Lamballais, E.}} \yr{2010}
  \at{Direct numerical simulation of a mixing layer downstream a thick splitter
  plate}.  \jt{Physics of Fluids}  \bvol{22}~(1),  \pg{015104}.

\bibitem[Lehmann {\em et~al.\/}(2014)Lehmann, Akins \&
  Little]{Lehmann&Little:AIAA2014}
{\sc \au{Lehmann, R.}, \au{Akins, D.} \& \au{Little, J.}} \yr{2014} Effects of
  ns-{DBD} plasma actuators on turbulent shear layers.  \bt{In {\em AIAA Paper
  2014-2220\/}}.

\bibitem[Little {\em et~al.\/}(2012)Little, Takashima, Nishihara, Adamovich \&
  Samimy]{Little&Samimy:AIAAJ2012}
{\sc \au{Little, J.}, \au{Takashima, K.}, \au{Nishihara, M.}, \au{Adamovich,
  I.~V.} \& \au{Samimy, M.}} \yr{2012}  \at{Separation control with
  nanosecond-pulse-driven dielectric barrier discharge plasma actuators}.
  \jt{AIAA Journal}  \bvol{50}~(2),  \pg{350--365}.

\bibitem[Mehta(1991)]{Mehta:EiF1991}
{\sc \au{Mehta, R.~D.}} \yr{1991}  \at{Effect of velocity ratio on plane mixing
  layer development: Influence of the splitter plate wake}.  \jt{Experiments in
  Fluids}  \bvol{10}~(4),  \pg{194--204}.

\bibitem[Monkewitz \& Huerre(1982)]{Monkewitz:PoF1982}
{\sc \au{Monkewitz, P.~A.} \& \au{Huerre, P.}} \yr{1982}  \at{Influence of the
  velocity ratio on the spatial instability of mixing layers}.  \jt{Physics of
  Fluids}  \bvol{25}~(7),  \pg{1137--1143}.

\bibitem[Nudnova {\em et~al.\/}(2010)Nudnova, Aleksandrov \&
  Starikovskii]{Nudnova:PPR2010}
{\sc \au{Nudnova, M.~M.}, \au{Aleksandrov, N.~L.} \& \au{Starikovskii, A.~Y.}}
  \yr{2010}  \at{Influence of the voltage polarity on the properties of a
  nanosecond surface barrier discharge in atmospheric-pressure air}.
  \jt{Plasma Physics Reports}  \bvol{36}~(1),  \pg{90--98}.

\bibitem[Papamoschou \& Roshko(1988)]{PapamoschouRoshko:JFM1988}
{\sc \au{Papamoschou, D.} \& \au{Roshko, A.}} \yr{1988}  \at{The compressible
  turbulent shear layer: an experimental study}.  \jt{Journal of Fluid
  Mechanics}  \bvol{197},  \pg{453--477}.

\bibitem[Popov(2011)]{Popov:JPD2011}
{\sc \au{Popov, N.~A.}} \yr{2011}  \at{Fast gas heating in a nitrogen--oxygen
  discharge plasma: I. kinetic mechanism}.  \jt{Journal of Physics D: Applied
  Physics}  \bvol{44}~(28),  \pg{285201}.

\bibitem[Post \& Corke(2004)]{PostCorke:AIAAJ2004}
{\sc \au{Post, M.~L.} \& \au{Corke, T.~C.}} \yr{2004}  \at{Separation control
  on high angle of attack airfoil using plasma actuators}.  \jt{AIAA Journal}
  \bvol{42}~(11),  \pg{2177--2184}.

\bibitem[Sabatini \& Bailly(2014)]{Sabatini:AIAAJ14}
{\sc \au{Sabatini, R.} \& \au{Bailly, C.}} \yr{2014}  \at{Numerical algorithm
  for computing acoustic and vortical spatial instability waves}.  \jt{AIAA
  Journal}  \bvol{53}~(3),  \pg{692--702}.

\bibitem[Samimy {\em et~al.\/}(2007)Samimy, Kim, Kastner, Adamovich \&
  Utkin]{Samimy:JFM2007}
{\sc \au{Samimy, M.}, \au{Kim, J.-H.}, \au{Kastner, J.}, \au{Adamovich, I.~V.}
  \& \au{Utkin, Y}} \yr{2007}  \at{Active control of high-speed and
  high-{R}eynolds-number jets using plasma actuators}.  \jt{Journal of Fluid
  Mechanics}  \bvol{578},  \pg{305--330}.

\bibitem[Seifert \& Pack(1999)]{SeifertPack:AIAAJ1999}
{\sc \au{Seifert, A.} \& \au{Pack, L.~G.}} \yr{1999}  \at{Oscillatory control
  of separation at high {R}eynolds numbers}.  \jt{AIAA Journal}  \bvol{37}~(9),
   \pg{1062--1071}.

\bibitem[Sharma {\em et~al.\/}(2011)Sharma, Bhaskaran \&
  Lele]{SharmaLele:AIAA2011}
{\sc \au{Sharma, A.}, \au{Bhaskaran, R.} \& \au{Lele, S.~K.}} \yr{2011}
  Large-eddy simulation of supersonic, turbulent mixing layers downstream of a
  splitter plate.  \bt{In {\em AIAA Paper 2011-208\/}}.

\bibitem[Sinha {\em et~al.\/}(2012)Sinha, Alkandry, Kearney-Fischer, Samimy \&
  Colonius]{SinhaSamimy:PoF2012}
{\sc \au{Sinha, A.}, \au{Alkandry, H.}, \au{Kearney-Fischer, M.}, \au{Samimy,
  M.} \& \au{Colonius, T.}} \yr{2012}  \at{The impulse response of a high-speed
  jet forced with localized arc filament plasma actuators}.  \jt{Physics of
  Fluids}  \bvol{24}~(12),  \pg{125104}.

\bibitem[Tian {\em et~al.\/}(2011)Tian, Ren, Xie, Wang, Zhou, Feng, Fu, Yang,
  Peng, Wang \& Liu]{Tian:ACSNANO11}
{\sc \au{Tian, H.}, \au{Ren, T.-L.}, \au{Xie, D.}, \au{Wang, Y.-F.}, \au{Zhou,
  C.-J.}, \au{Feng, T.-T.}, \au{Fu, D.}, \au{Yang, Y.}, \au{Peng, P.-G.},
  \au{Wang, L.-G.} \& \au{Liu, L.-T.}} \yr{2011}  \at{Graphene-on-paper sound
  source devices}.  \jt{ACS Nano}  \bvol{5}~(6),  \pg{4878--4885}.

\bibitem[Vukasinovic {\em et~al.\/}(2010)Vukasinovic, Rusak \&
  Glezer]{VukasinovicGlazer:JFM2010}
{\sc \au{Vukasinovic, B.}, \au{Rusak, Z.} \& \au{Glezer, A}} \yr{2010}
  \at{Dissipative small-scale actuation of a turbulent shear layer}.
  \jt{Journal of Fluid Mechanics}  \bvol{656},  \pg{51--81}.

\bibitem[Wiltse \& Glezer(1998)]{Wiltse:PF98}
{\sc \au{Wiltse, J.~M.} \& \au{Glezer, A.}} \yr{1998}  \at{Direct excitation of
  small-scale motions in free shear flows}.  \jt{Physics of Fluids}
  \bvol{10}~(8),  \pg{2026--2036}.

\bibitem[Winant \& Browand(1974)]{Winant74}
{\sc \au{Winant, C.~D.} \& \au{Browand, F.~K.}} \yr{1974}  \at{Vortex pairing:
  the mechanism of turbulent mixing-layer growth at moderate {R}eynolds
  number}.  \jt{Journal of Fluid Mechanics}  \bvol{63}~(02),  \pg{237--255}.

\bibitem[Wu \& Wu(1993)]{WuWu:JFM1993}
{\sc \au{Wu, J.-Z.} \& \au{Wu, J.-M.}} \yr{1993}  \at{Interactions between a
  solid surface and a viscous compressible flow field}.  \jt{Journal of Fluid
  Mechanics}  \bvol{254},  \pg{183--211}.

\bibitem[Yeh {\em et~al.\/}(2017)Yeh, Munday \& Taira]{Yeh:AIAA2017}
{\sc \au{Yeh, C.-A.}, \au{Munday, P.} \& \au{Taira, K.}} \yr{2017} Use of local
  periodic heating for separation control on a {NACA} 0012 airfoil.  \bt{In
  {\em AIAA Paper 2017-1451\/}}.

\bibitem[Zhuang \& Dimotakis(1995)]{ZhuangDimotakis:PoF1995}
{\sc \au{Zhuang, M.} \& \au{Dimotakis, P.~E.}} \yr{1995}  \at{Instability of
  wake-dominated compressible mixing layers}.  \jt{Physics of Fluids}
  \bvol{7}~(10),  \pg{2489--2495}.

\end{thebibliography}

\end{document}